\documentclass[aps,pra,epsf,superscriptaddress,amsmath,amssymb,amsfonts,twocolumn,showpacs,nofootinbib,10pt]{revtex4-1}

\usepackage{graphicx}
\usepackage{epsfig}
\usepackage{dcolumn}
\usepackage{bm}
\usepackage{braket}
\usepackage{amsmath}
\usepackage{mathtools}
\usepackage{graphicx,color,xcolor}
\usepackage{hyperref}
\newcommand{\abs}[1]{\left| #1 \right|} 
\usepackage{hyperref}
\usepackage{multirow}
\allowdisplaybreaks

\usepackage[normalem]{ulem} %% REMOVE ME!

\begin{document}

\title{Stability and mixed phases of three-component droplets in one dimension}

\author{I. A. Englezos}
\email{ilias.englezos@uni-hamburg.de}
\affiliation{Center for Optical Quantum Technologies, Department of Physics, University of Hamburg, 
Luruper Chaussee 149, 22761 Hamburg Germany}
\author{E.~G.~Charalampidis}
\email{echaralampidis@sdsu.edu}
\affiliation{Department of Mathematics and Statistics and Computational
Science Research Center, San Diego State University, San Diego, CA 92182-7720, USA}
\author{P. Schmelcher}
\affiliation{Center for Optical Quantum Technologies, Department of Physics, University of Hamburg, 
Luruper Chaussee 149, 22761 Hamburg Germany} 
%\affiliation{The Hamburg Centre for Ultrafast Imaging, University of Hamburg, Luruper Chaussee 149, 22761 Hamburg, Germany} 
\author{S. I. Mistakidis}
\affiliation{Department of Physics, Missouri University of Science and Technology, Rolla, MO 65409, USA}

\date{\today}

\begin{abstract} 

We explore the ground state properties and excitation spectra of one-dimensional, three-component bosonic mixtures, where two majority components form a self-bound droplet coupled to a third, minority component. 
Within a self-consistent Lee-Huang-Yang framework, we identify distinct mixed self-bound droplet phases and characterize their phase transitions through variations of either the particle number of the majority components or the intercomponent coupling. 
The ensuing phases demonstrate that the minority component is untrapped, partially trapped, or fully trapped by the majority droplet species.
These states are characterized by their binding energies captured by the chemical potentials and  their low-amplitude excitation spectrum, including mode crossings at the particle-emission threshold. 
We further derive effective reduced models in the high-imbalance limit, which accurately reproduce the numerically computed ground states, while providing analytical insights into the role of quantum fluctuations.
Our results establish the stability and structure of mixed droplet phases provide experimentally relevant signatures for their identification.
\end{abstract}

\maketitle

\section{Introduction}

Quantum droplets are self-bound, ultradilute, liquid-like many-body states, whose formation entails the presence of quantum fluctuations at weakly interacting macroscopic ultracold settings~\cite{PfauReview,MalomedReview,MalomedLuoReview,mistakidis2023few}.
In this context, the first-order Lee-Huang-Yang (LHY)~\cite{LeeHuangYang1957} quantum correction term to the mean-field (MF) energy is often sufficient to describe the emergence of droplet states~\cite{Petrov2015, PetrovLowD, Englezos2025Droplets}. 
These exotic phases of matter were first experimentally observed in long-range dipolar bosonic gases~\cite{KadauDropExp,Chomaz2016,PfauReview,Chomaz_2023}, and soon thereafter, in two-component short-range interacting bosonic mixtures~\cite{CabreraTarruellDropExp,CheineyTarruellDropExp,SemeghiniFattoriDropExp,FortHeteroExp,GuoHeteroDrop2021,Cavicchioli}. 
In these latter settings, in which we also focus herein, it is important to determine the species' density ratio since i) if it is fixed (and dictated by the intracomponent interaction fraction), the components behave identically, thus yielding a symmetric droplet with enhanced stability~\cite{Petrov2015}, while ii) if it is broken, genuine two-component droplet configurations arise~\cite{FlynnPRR2023,FlynnTrapped2023,QuantumCrit2023,Pelayo_2025}.

Recently, a large number of studies on symmetric droplet setups in different dimensionalities have explored, among others, their ground states and spectrum~\cite{AstrakharchikMalomed1DDynamics,Collective1D,Mistakidis2021,Englezos2023,zhao2025bulk} (including clusters  thereof~\cite{dong2026robust}), inelastic collisions~\cite{FattoriCollisions,AstrakharchikMalomed1DDynamics}, as well as their ability to host solitons~\cite{Katsimiga_solitons,Edmonds_solitons}, vortices~\cite{Li_vortex,Bougas_vortex,Tengstrand_vortex,Yoifmmode}, dispersive shock waves~\cite{Chandramouli}, kinks~\cite{kartashov2022spinor,katsimiga2023interactions,Mistakidis_kink} and rogue waves~\cite{chandramouli2025rogue}.
By contrast, genuinely two-component droplets that break the aforementioned density-ratio condition, e.g. by considering particle~\cite{Englezos2024}, mass~\cite{FortHeteroExp,Mistakidis2021} or interaction~\cite{gangwar2025interaction} imbalance between the two-components, are arguably far less investigated.
A handful of current results revealed mixed droplet-gas phases~\cite{QuantumCrit2023,FlynnPRR2023,FlynnTrapped2023,Pelayo_2025}, the existence of excited multipole droplet states~\cite{Kartashov_multipole}, and droplets of a super-Tonks gas confined in an optical lattice~\cite{latticeDrop2023}, as well as the droplet stability in a ring geometry~\cite{ReimannRotatingAndVortex} and in a  harmonic trap~\cite{charalampidis2024two}, see also Refs.~\cite{Englezos2024,Boudjemaa_2023,ParisiGiorginiMonteCarlo,ParisiMonteCarlo2019}.   
A challenge in studying particle-imbalanced two-component droplets is the significantly different length scales of the involved species, which has been typically circumvented by the inclusion of external traps, which, however, formally fall outside the validity of the original LHY theory.

A conceivable approach to mitigate such constraints is to embed a minority species within a droplet. 
This would allow to indirectly limit the droplet extent via tuning the droplet-minority species attraction as it was, for instance, argued recently using attractive potential wells in a droplet~\cite{debnath2023interaction,bristy2025localization} or by considering a third impurity-component characterized by weak  intracomponent interactions~\cite{Boudjemaa_2020,Diplaris_2026}. 
Such an approach can be, among others, practical for forthcoming experimental realizations. 
Moreover, the crossover from the impurity limit to particle balance between the third component and the droplet is yet to be explored. 
Additionally, several important questions remain unresolved, such as i) can the third component act as a knob for the droplet structure~\cite{jogania2025oscillation}?, ii) what are the phases of the underlying three-component configurations?, and iii) are there parameter regimes where the minority species is fully or partially trapped by the droplet~\cite{ma2021borromean,Edler2022bubbles,Cui2025_3CompDroplets}?
In this vein, there are even broader implications of such a setup which also provides links with impurity physics~\cite{Bighin_impurity,Pelayo_BFdrop,Sinha2023}, a topic that has been extensively  addressed in the past decade with repulsive Bose gases~\cite{BosePolaronDemler,scazza2022repulsive,mistakidis2023few,PolaronsDemler}, but it is currently unexplored in the droplet realm. 

Motivated by these open questions, we explore the droplet phases and their stability in a three-component bosonic mixture in one dimension (1D) in this work. 
The lower dimensionality is expected to facilitate the prolongation of droplet lifetimes due to the relatively lower densities~\cite{BourdelCrossover}.
Our description relies on a fully self-consistent LHY framework yielding the suitable coupled  extended Gross-Pitaevskii equations (eGPEs) derived for the 1D mixture in~\cite{Englezos2025Droplets}, see also Refs.~\cite{ma2021borromean, Cui2025_3CompDroplets} for generalizations to three dimensions. 
In contrast to previous treatments, this framework is valid across the MF stability regime, and accounts for the LHY term of the third component facilitating its stability. 

We focus on homonuclear systems where the two components are equivalent (in terms of their mass, atom number and interaction parameters) forming a strongly self-bound symmetric  droplet that is coupled to a third, minority component.  
Unlike the impurity limit addressed in Ref.~\cite{Diplaris_2026}, here the third species containing a finite particle number actively participates in the formation of the self-bound three-component state. 
A key result of our work is the emergence of distinct mixed droplet phases where the minority component is untrapped, partially trapped, or fully trapped by the symmetric  droplet upon considering variations of either the intercomponent interaction strength or the particle imbalance between the droplet and the third component. 
These phases do not share a two-component counterpart and are characterized through their chemical potentials (being negative throughout the considered parametric variations), the intercomponent spatial overlap, and low-amplitude excitation spectra.

The latter are obtained by constructing a Bogoliubov-de Gennes (BdG) framework for the three-component system, which constitutes an additional important aspect of our study and extends previous works on symmetric~\cite{Collective1D} and two-component~\cite{charalampidis2024two} droplets. 
Notably, the ensuing spectra reveal that in the transition from untrapped to fully trapped minority species, the collective-mode branches cross the particle emission threshold from below as the intercomponent attraction increases.
Additionally, other intriguing spectral fingerprints are found such as mode pairing and avoided crossings associated with the aforementioned transition. 
These spectral signatures which are inherently related to the three-component settings but do not have two-component counterparts are expected to assist in future studies aiming to controllably excite droplet modes where the minority species can act as a ``sensor". 
Finally, we derive effective reduced models in the limit of large particle imbalance. 
They provide analytical insights for the structure of the components and quantitatively reproduce the  numerical results of the full eGPEs for weak and moderate intercomponent couplings, while deviating for strong attractions.

This work is structured as follows. 
In Section~\ref{sec:ThreeComponent}, we introduce the three-component mixture under investigation governed by the suitable eGPEs, while briefly discussing the BdG framework that is leveraged to extract the low-amplitude excitation spectrum of the system.
In Section~\ref{sec:results}, we address the ground state configurations and excitation spectrum of both particle balanced and particle imbalanced settings for varying atom number and intercomponent coupling strengths.
Section~\ref{sec:ParticleImbalance} provides analytical insights for the interpretation of the three-component structures by constructing relevant effective reduced models based on LHY theory in the limit of large particle imbalance. 
We summarize and propose extensions of our results for future research in Section~\ref{sec:SummaryAndOutlook}.
Appendix~\ref{app:SM} elaborates on the derivation of the BdG equations,  and Appendix~\ref{app:limit} outlines the calculation of the approximate LHY energy utilized in the reduced models. 
Appendix~\ref{app:MF} elucidates the impact of the LHY term of the minority component to feature self-bound structures.
Finally, Appendix~\ref{app:phase_diagram} elaborates on the impact of intercomponent interaction strengths to the ground state energy of the three-component setting.

\section{Three-component Homonuclear Droplet setup}\label{sec:ThreeComponent}

By construction, three-component (nonlinearly interacting) mixtures offer additional possibilities for intercomponent asymmetry that could lead to enriched phases of matter and intriguing  nonequilibrium phenomena which do not exist in their two-component counterparts.
However, exploring their behavior in the full parameter space containing at least twelve independent parameters (namely six interaction strengths, three masses, and three atom numbers) is a formidable task. 
As such, it is natural to first study these systems by considering specific symmetries which reduce the number of the aforementioned free parameters. 
A physically motivated setting that respects this requirement is a homonuclear mixture, where $m_A=m_B=m_C \equiv m$. 
Additionally, it is viable to envisage that the individual components experience (almost) the same intracomponent interactions, i.e., $g_A=g_B=g_C \equiv g$, a scenario that is in principle possible using Feshbach resonances~\cite{chin2010feshbach}, and that two of them accommodate the same atom number, $N_A=N_B \equiv N$. 
The latter can be accomplished by employing radiofrequency transitions among the involved hyperfine states. 
Hence, two of the components (denoted here as $A$ and $B$) are equivalent, whereas the third one (called $C$) is different which also implies that $g_{AC}=g_{BC} \equiv G_C$. 
Accordingly, the number of free parameters is reduced to six (i.e., $g$, $g_{AB}$, $G_{C}$, $m$, $N$ and $N_C$) of which $g$ and $m$ are held fixed since they correspond to a specific atomic element. 

These assumptions in the interaction regime of repulsive intracomponent ($g>0$) and attractive intercomponent ($g_{AB}<0$) couplings favor the formation of a so-called symmetric droplet~\cite{Petrov2015} in species $A$, $B$, while component $C$ is independent, see also Ref.~\cite{ma2021borromean} for a similar 3D system. 
Interestingly, the aforementioned composition allows to use the third component as a ``knob" for configuring unseen complex droplet phases of matter. 
A candidate for realizing such a system in near term experiments consists, for instance, of  three $^{39}$K  hyperfine states two of which have already been utilized in corresponding 3D droplet experiments~\cite{CheineyTarruellDropExp,CabreraTarruellDropExp}. 
Additionally, without loss of generality, we consider a 1D box confinement of typical length $L=300$ (being sufficiently extended to not impact the ensuing droplet configurations) along the elongated $x$-direction, while the transverse ones ($y,z$) are assumed to be tightly confined characterized by a frequency $\omega_{\perp}$. 
This, together with the requirement $\abs{\mu+\mu_C} \ll \hbar \omega_{\perp}$, with $\mu$ ($\mu_C$) representing the chemical potential of the droplet (third) component, ensures that the atoms are kinematically constrained across the $x$-direction. 
For completeness, we remark here that 1D droplet experiments are still lacking.

\subsection{Droplet coupled to another component}

Our many-body system comprises a bosonic droplet assembled by components $A$, $B$ and a third component $C$ interacting with the symmetric two-component droplet via $G_C$. 
Recently, it was analytically demonstrated~\cite{Englezos2025Droplets} that the LHY energy density of such a  
three-component mixture under the above-described symmetry conditions takes the form  
\begin{equation}
\begin{split}
\label{ELHY2Plus1}
\mathcal{E}_{\textrm{LHY}} =& - \frac{\sqrt{2 m}}{6\hbar\pi} \Big( (2 (g-g_{AB})n)^{\frac{3}{2}}\\ 
& +  (W + Q)^{\frac{3}{2}} + (W - Q)^{\frac{3}{2}}   \Big),
\end{split}
\end{equation}
with the parameters  
\begin{subequations}\label{ELHY2Plus1Params}
\begin{align}
Q =& \Big( [(g + g_{AB}) n - g_{C} n_C ]^2 + 8 G_C^{2} n n_C  \Big)^{\frac{1}{2}},\\
W =&\, (g + g_{AB}) n + g_{C} n_C.
\end{align}
\end{subequations} 
In these expressions, $n(x)$ and $n_C(x)$ are the densities of the two symmetric components ($A$, $B$) and of component $C$. 
The LHY energy density is real and negative admitting droplet solutions as long as $g>g_{AB}$ and $W - Q > 0 \xrightarrow{} 2 G_C^2 < g_{C}(g + g_{AB})$, while it becomes complex otherwise. 
The interaction interval $2 G_C^2 < g_{C}(g + g_{AB})$ defines the MF stability regime of the three-component mixture as it was argued in~\cite{Englezos2025Droplets}. 
Note also that in the absence of the third component, i.e. when $G_C=g_{C}=n_C=0$, Eq.~(\ref{ELHY2Plus1}) reduces to the LHY energy of a symmetric droplet~\cite{Englezos2025Droplets,PetrovLowD}.

The corresponding coupled set of 1D eGPEs describing the dynamics of the three-component droplet setting with $\psi_A= \psi_B \equiv \psi$ reads~\cite{Englezos2025Droplets}  
\begin{subequations}
\label{eGPETwoPlusOneComponent}
\begin{align}
\label{eGPETwoPlusOneComponent_a}
2i\hbar\frac{\partial \psi}{\partial t}& = - \frac{\hbar^2}{m} \frac{\partial^2 \psi}{\partial x^2} +\Big( 2(g+g_{AB}) \abs{\psi}^2 + \nonumber \\
&+ 2G_C \abs{\psi_C}^2 +\frac{\partial \mathcal{E}_{\textrm{LHY}}}{\partial n} \Big)\psi,  \\ 
i\hbar\frac{\partial \psi_C}{\partial t}& = - \frac{\hbar^2}{2m} \frac{\partial^2 \psi_C}{\partial x^2} + \Big( g_{C}\abs{\psi_{C}}^2  + \nonumber \\
&+ 2G_C\abs{\psi}^2 + \frac{\partial \mathcal{E}_{\textrm{LHY}}}{\partial n_C} \Big)\psi_C. \label{eGPETwoPlusOneComponent_b}
\end{align}
\end{subequations}
Here, $n(x)=\abs{\psi(x)}^2$ and $n_C(x)=\abs{\psi_C(x)}^2$. 
Importantly, these eGPEs contain the LHY terms of both the symmetric  droplet and the third component, as well as an intercomponent LHY contribution. 
This is in sharp contrast to previous three-component considerations, in which only MF couplings were taken into account between the droplet and the other component~\cite{abdullaev2020bosonic}. 
It is also important to note that the LHY term appearing in the equation of the third component [Eq.~(\ref{eGPETwoPlusOneComponent_b})] is responsible for the stabilization of component $C$ into a self-bound configuration, see also Appendix~\ref{app:MF} for a detailed analysis. 
Hence, its contribution is crucial for describing three-component self-bound states. 
This aspect, to the best of our knowledge, has not been discussed in previous studies. 

Moreover, the partial derivatives of the LHY energy density entering the above equations of motion~(\ref{eGPETwoPlusOneComponent_a})-(\ref{eGPETwoPlusOneComponent_b}) can be readily derived from Eq.~(\ref{ELHY2Plus1}), and acquire the form  
\begin{subequations} 
\begin{align}
\label{LHYTerms3Comp}
\frac{\partial\mathcal{E}_{\textrm{LHY}}}{\partial n} =& - \frac{\sqrt{2m}}{4\pi \hbar  } \Bigg(  \Big( g + g_{AB} + \frac{Z}{Q} \Big) \sqrt{W - Q} \nonumber \\
&+ \Big(  g + g_{AB} - \frac{Z}{Q}\Big) \sqrt{W + Q} \nonumber \\
&+ 2 \sqrt{2}  (g - g_{AB})^{\frac{3}{2}} \sqrt{n}    \Bigg), \\
\frac{\partial\mathcal{E}_{\textrm{LHY}}}{\partial n_C} =& - \frac{\sqrt{2m}}{4\pi \hbar }  \Bigg(  \Big(  g_C + \frac{P}{Q} \Big)  \sqrt{W - Q} \nonumber \\
&+  \Big( g_C - \frac{P}{Q} \Big) \sqrt{W + Q}   \Bigg),
\end{align}
\end{subequations}
with
\begin{subequations} 
\begin{align}
\label{2Plus1DerParams}
Z = & -(g+g_{AB})^2n + (g_C(g+g_{AB}) - 4 G_C^{2} )n_C, \\
P = & (g_{C} (g+g_{AB}) - 4 G_C^{2}) n - g_{C}^{2} n_C.
\end{align}
\end{subequations}
In what follows, the energy scale is set by $m g^2/\hbar^2$.  
Hence, the length, time, and interaction strengths are expressed in units of $\hbar^2/(m g)$, $\hbar^3/(m g^2)$, and $g$ respectively.

To identify stationary droplet solutions of our three-component system, we consider the stationary version of Eqs.~(\ref{eGPETwoPlusOneComponent_a})-(\ref{eGPETwoPlusOneComponent_b})
that is obtained by the substitution of the (separation of variables) ansatz: $\psi_j
(x,t)=\Psi_j(x)e^{-i \mu_j t}$ therein. 
The field $\Psi_j(x)$ is the $j$th steady state to be obtained, and $\mu_j$ represents its corresponding chemical potential. 
We note that for the two symmetric  components that we will be considering below, $\mu_A=\mu_B \equiv \mu$ holds. 
To identify the steady states  $\Psi_j(x)$ numerically, we introduce a one-dimensional computational domain $[-L,L]$ with $L=300$ that is supplemented with homogeneous Dirichlet boundary conditions at its endpoints. 
The spatial derivatives that arise in the steady-state problem are replaced by a 4th-order accurate, centered finite difference formula. 
Then, we employ Newton's method (and variants thereof)~\cite{kelley2003solving} to solve the underlying coupled nonlinear system of algebraic equations.

We would like to note in passing that the present setup is defined on the free space, i.e., in the absence of external potential (unlike our recent work for a two-component system in~\cite{charalampidis2024two}). 
That by itself, introduces a non-trivial kernel in the underlying Jacobian matrix that is associated with the invariances of the system. 
Specifically the (right) nullspace vectors therein correspond to the phase invariance and translation invariance (again, under the absence of an $x$-dependent potential) of the system. 
We restore robust convergence in Newton's method by augmenting the system through the introduction of suitable phase conditions and Lagrange multipliers that help factor out the degeneracies. 
For further details about this approach and the implementation of phase conditions, we refer the reader to Ref.~\cite{uecker_book}.

\subsection{BdG Equations and Stability Matrix}\label{sec:BdG_stab_matrix}

To assess the stability of the computed stationary three-component droplet solutions, we deploy the BdG formalism. 
In particular, having at hand the stationary solutions, $\Psi(x)$, $\Psi_C(x)$ computed using Newton's method at the respective chemical potentials $\mu$, and $\mu_C$, we introduce the small amplitude $\epsilon \ll 1$ perturbation Ans\"atze: 
\begin{subequations}
\label{perturbation}
\begin{align}
\Psi^{{\rm BdG}} &= e^{-i\mu t} [ \Psi + \epsilon (a e^{i\omega t} + b^* e^{-i\omega^* t}) ], \\
\Psi^{{\rm BdG}}_C &= e^{-i\mu_C t} [ \Psi_C + \epsilon (a_C e^{i\omega t} + b_C^* e^{-i\omega^* t}) ], 
\end{align}
\end{subequations}
for the eGPEs of Eqs.~(\ref{eGPETwoPlusOneComponent_a})-(\ref{eGPETwoPlusOneComponent_b}).
This yields the underlying linearized (i.e., of order $\mathcal{O}(\epsilon)$) system of the  equations of motion. 
Here, $a=a(x)$, $b=b(x)$, $a_C=a_C(x)$, $b_C=b_C(x)$ are spatially dependent complex functions composing the eigenvectors $[a(x), b(x)]^{\intercal}$ and $[a_C(x), b_C(x)]^{\intercal}$, while $\omega$ represents the eigenfrequency.  

We refrain from presenting these calculations here, as they are rather lenghty.
After separating the terms $\propto e^{i\omega t}$ and $\propto e^{-i\omega^* t}$, we arrive at a coupled system of four eigenvalue equations for the four eigenvector components and the eigenfrequency $\omega$.
Defining the vector $v=(a, b, a_C, b_C)$, the resulting equations can be expressed in matrix form as ${\rm diag}(-1, 1, -1, 1) \mathcal{A} v = \omega  v $. 
The derivation of the elements of the $4 \times 4$ stability matrix $\mathcal{A}$ is given in Appendix~\ref{app:SM}. 
By solving the resulting eigenvalue problem, we extract the energy spectrum containing the ensuing eigenmodes of the elementary small amplitude excitations of the stationary solutions $\Psi(x)$ and $\Psi_C(x)$.
Complex eigenvalues indicate that the corresponding solution is dynamically unstable, while purely real eigenvalues (i.e., with zero imaginary part) indicate stable configurations~\cite{keith_book}.
~The real part of the spectrum encompasses the collective excitation branches of a given configuration.

\section{Mixed droplet phases}\label{sec:results}

Below, we focus on the case of a strongly attractive intercomponent interaction (i.e., close to the MF stability edge) between the two symmetric components, namely $g_{AB}=-0.96g$. 
This ensures that these components assemble into a tightly self-bound single-component droplet~\cite{PetrovLowD} in the absence of the third component. 
Moreover, to reduce the free parameter space of our setting, we fix the particle number of the third independent component $C$ to $N_C=10$ and its intracomponent coupling to $g_C=g$. 
Under these conditions, the MF stability regime is given by $\abs{G_C} < 0.1 \sqrt{2} \approx 0.14$.
Our analysis is based on the three-component droplet ground state densities to infer the emergent spatial configurations along with the underlying excitation spectrum explicating their stability and behavior of collective modes. 
Additionally, the chemical potential allows to deduce the bound state nature of the three-component droplet structures. 
We address both particle-balanced and particle-imbalanced regimes between the symmetric droplet (species $A$, $B$ species) and the third component ($C$), allowing us to explore the partial or full trapping of the component $C$ by the droplet.

Let us explicitly note here that for $G_C=0$ (dubbed decoupled limit, see also Sec.~\ref{sec:ParticleImbalance}) the three-component setting decouples into two independent droplet subsystems, namely components $A$ and $B$ constitute the first and component $C$ the second\footnote{It can be shown that for $G_C=0$, the minority component $C$ satisfies the single-component droplet equation $i\hbar\frac{\partial \Psi_C}{\partial t} = \Big( - \frac{\hbar^2}{2m} \frac{\partial^2}{\partial x^2} + g_C\abs{\Psi_C}^2 
-\frac{\sqrt{mg_C^3}}{\hbar \pi} \abs{\Psi_C} \Big) \Psi_C $.}. 
At $G_C=0$, the symmetric two-component droplet is characterized by saturation density $n^s = \frac{mg^3 ( 1 + 3(g_{AB}/g)^2 + [1-(g_{AB}/g)^2]^{3/2} ) }{9\hbar^2\pi^2(g+g_{AB})^2} \approx 53.3 $ with saturation chemical potential $\mu^s = -n^s(g+g_{AB})/2\approx - 1.066$, while the (single-component) droplet of component $C$  features saturation density $n^s_C = \frac{4mg_C }{9\hbar^2\pi^2}\approx 0.045$ and saturation chemical potential $\mu_C^s = -g_Cn_C^s/2 \approx - 0.0225$~\cite{Englezos2025Droplets}.

\begin{figure}[ht]
\centering   \includegraphics[width=1.02\linewidth]{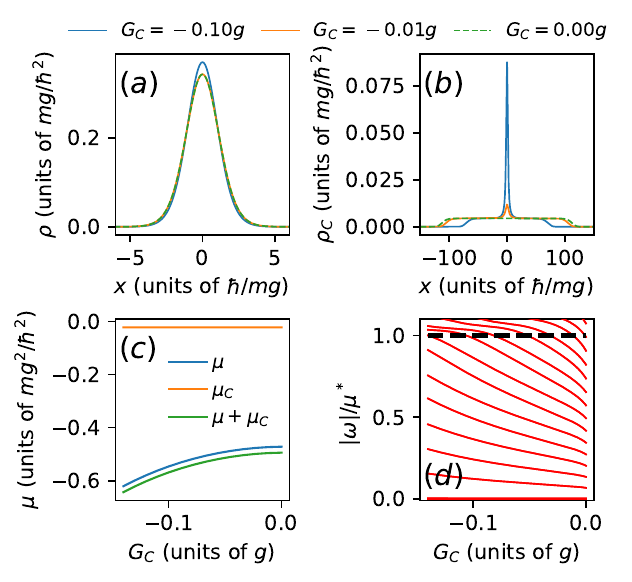}
\caption{
Overview of particle-balanced, three-component droplet configurations. 
Ground-state densities per particle of (a) the symmetric two-component  droplet and (b) the third independent component $C$ for varying intercomponent coupling $G_C$ (see legend). 
The third component is either untrapped or partially trapped by the droplet for $G_C=0$ and $G_C<0$ respectively. 
(c) Chemical potentials of the symmetric droplet ($\mu$), the $C$ component ($\mu_C$) and their sum $\mu+\mu_C$ as a function of $G_C$. 
All chemical potentials are negative, indicating that the corresponding configurations are strongly bound, with stronger binding for increasingly negative $G_{C}$.
(d) Elementary excitation spectrum of the particle-balanced configurations. 
The horizontal thick dashed line corresponds to the particle emission threshold, $\mu^*= \min(\abs{\mu},\abs{\mu_C})$. 
Evidently, component $C$ remains only partially trapped by the two symmetric  components for all values of $G_C$, with the untrapped fraction forming an extended single-component FT droplet. 
The particle-balanced three-component mixture consists of $N=N_C=10$ atoms, with $g_C=g$, and $g_{AB}=-0.96g$.}
\label{fig:N10Results}
\end{figure}

\subsection{Particle-balanced configurations: partial trapping of the third component}\label{ParticleBalanced}

The ground-state density distributions of the symmetric  droplet, $\rho(x)=\abs{\Psi}^2/N$, consisting of two identical species and the third component, $\rho_C(x)=\abs{\Psi_C}^2/N_C$, are presented in Fig.~\ref{fig:N10Results}(a), (b) for fixed $N=N_C=10$ and different intercomponent interactions $G_C \leq 0$.   
As it can be seen, in the decoupled limit where $G_C=0$ the droplet and species $C$ are independent. 
The former assembles in a localized Gaussian-like configuration and the latter forms a flat-top (FT)  structure. 
This situation implies that component $C$ is untrapped by the droplet. 
Turning our focus into attractive $G_C$, we observe that $\rho(x)$ maintains the Gaussian droplet profile regardless of the value of $G_C$, and it exhibits a slight localization tendency for larger attractions accompanied by a small amplitude density hump at $x=0$ (hardly visible in Fig.~\ref{fig:N10Results}(a)).
This means that the back-action of the comparatively delocalized component $C$ to the droplet is suppressed, owing to their relatively weak coupling $\abs{G_C} \ll \abs{g_{AB}}$.

In sharp contrast, component $C$ undergoes significant structural deformations upon tuning $G_C<0$  as illustrated in Fig.~\ref{fig:N10Results}(b). 
Namely, for relatively small attractive $G_C$, a density peak appears in $\rho_C(x)$ which simultaneously shows a localization trend. 
This behavior implies that a fraction of atoms from component $C$ starts to accumulate within the symmetric droplet.   
For stronger attractive $G_C$, the aforementioned density hump of $\rho_C(x)$ around $x=0$  becomes significantly enhanced, and $\rho_C(x)$ features a noticeable localization since a larger amount of $C$ atoms are attracted, and in particular, trapped by the droplet. 
Here, it is evident that the atom fraction of species $C$ trapped by the droplet is relatively small as can also be inferred from the somewhat small amplitude of the peak of $\rho_C(x)$ as compared to the peak of $\rho(x)$ which explains the minor impact of the former to the latter.

In all cases, however, component $C$ is only partially trapped by the symmetric droplet, while the remaining atoms reside outside the droplet, forming a single-component FT structure in $\rho_C(x)$ [Fig.~\ref{fig:N10Results}(b)]. 
This FT structure has a substantially lower (higher) density (spatial extent) when compared to the droplet made by the two symmetric components and to the density peak of component $C$.  
Moreover, the FT density  of component $C$ remains constant at the single-component FT droplet density predicted in~\cite{PetrovLowD}, i.e. $\rho_C^s=n_C^s/N_C\approx 0.0045$, irrespective of $G_C$, see also the discussion in Sec.~\ref{sec:ParticleImbalance} for relevant analytical arguments of this effect.  

To confirm the bound state nature of the studied configurations, we compute the chemical potentials of the symmetric droplet $\mu$ and the third component $\mu_C$ as well as the total chemical potential $\mu_{\rm tot}=\mu+\mu_C$ next. 
These are shown in Fig.~\ref{fig:N10Results}(c) as a function of the intercomponent coupling $G_C$.
We note that the chemical potentials were computed by considering the total number of atoms as our principal bifurcation/continuation parameter, following our approach in~Ref.~\cite{charalampidis2024two}.
We find that both $\mu<0$ and $\mu_C<0$, thus verifying that all components are self-bound even at $G_C \to 0$. 
Specifically, $\mu_C$ exhibits a slight decreasing trend (hardly visible in Fig.~\ref{fig:N10Results}(c)) but overall remains nearly constant close to its saturation value $\mu_C^s = -\frac{2mg_C^2}{9\hbar^2\pi^2} \approx -0.0225$~\cite{Englezos2025Droplets}. 
On the other hand, $\mu$ acquires significantly more negative values compared to $\mu_C$, which is attributed to the strong interaction ($g_{AB}=-0.96g$) between the components $A$ and $B$ forming a droplet even in the absence of component $C$. 
Furthermore, as expected, $\mu$ experiences a noticeable decreasing trend for increasing intercomponent attraction $|G_{C}|$ due to the enhanced binding with component $C$. 
Consequently, the total chemical potential, $\mu_{\rm {tot}}=\mu+\mu_C$, follows the same trend as $\mu$. 
The fact that $\abs{\mu} < \abs{\mu_{\rm {tot}}}$ indicates a stronger bound three-component configuration compared to the (two-component) majority droplet, highlighting the role of the third component in enhancing the overall binding beyond that of conventional two-component droplet configurations~\cite{PetrovLowD}.

\begin{figure*}[ht]
\centering   
\includegraphics[width=0.95\linewidth]{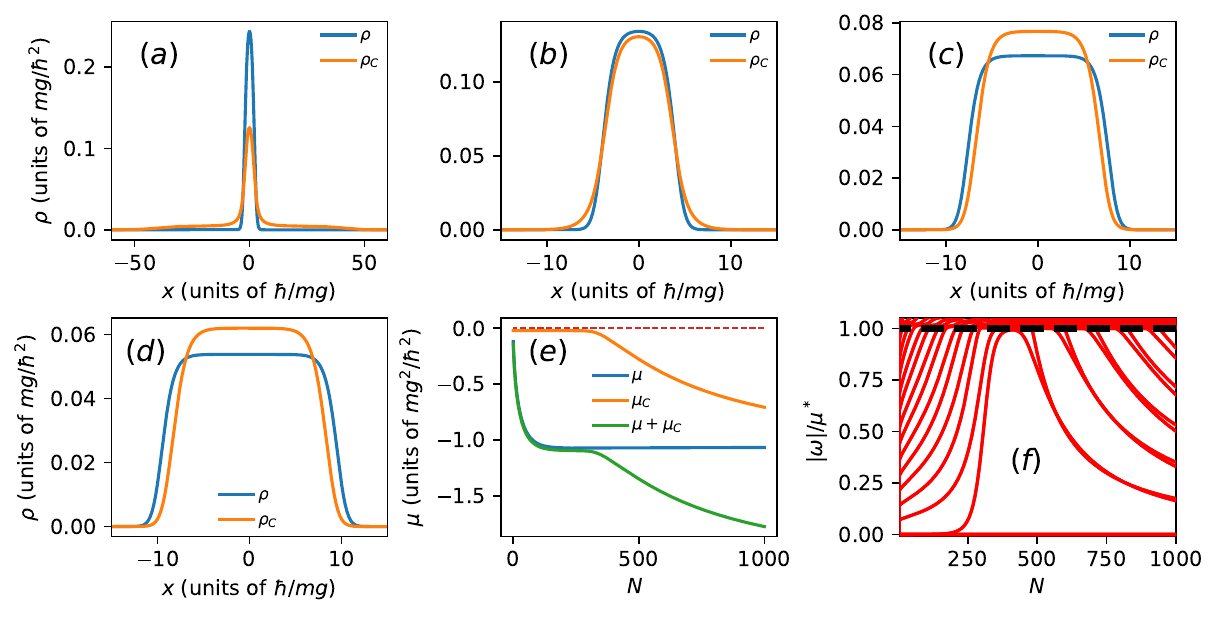}
\caption{Distributions of a particle imbalanced three-component mixture with $N_C=10$ minority atoms in component $C$, fixed interactions $g=g_C$, $g_{AB}=-0.96g$, and $G_C=-0.01g$, and varying particle number $N$ of the symmetric droplet. 
Ground-state densities of the individual components (see legends) for (a) $N=200$, (b) $N=400$, (c) $N=800$ and (d) $N=10^3$. 
Two distinct phases representing partial (panels (a), (b)) or full trapping (panels (c), (d)) of component $C$ by the symmetric  droplet can be discerned.
(e) The total chemical potential and the ones of the symmetric droplet and component $C$ (see legend) with respect to $N$. 
The horizontal dashed red line marks $\mu=0$. 
A notable decrease of $\mu_C$ in the case of full trapping occurs. 
(f) Excitation spectrum of the three-component mixture for varying $N$.  
Evidently, upon increasing $N$, the energy of the elementary excitations rises and eventually crosses the particle emission threshold (thick dashed line) as component $C$ becomes fully trapped by the (two-component) droplet. 
Afterwards, increasing $N$ further yields the particle emission threshold crossing of different excitations from above.
}
\label{fig:ContinuationOverN}
\end{figure*}

Finally, we examine the excitation spectrum of the above-discussed three-component structures to understand their stability and collective mode behavior, see Fig.~\ref{fig:N10Results}(d). 
As outlined in Sec.~\ref{sec:BdG_stab_matrix}, the excitation spectrum is obtained using the BdG analysis, see also Appendix~\ref{app:SM} for the matrix elements of the linearized equations. 
It turns out that the imaginary parts of the eigenvalues for all configurations studied throughout this work are zero as expected since we focus our investigations on ground states. 
Meanwhile, the real part of the spectrum is rich containing a plethora of elementary excitations that are energetically equidistant at $G_C=0$, and their energy gap increases as the intercomponent coupling $G_C$ becomes more attractive. 
The aforementioned increasing energy gap is associated with the interaction-dependent behavior of the collective modes whose branches exhibit an increasing tendency for larger attractive $G_C$. 
Interestingly, the higher-lying collective modes cross the corresponding particle emission threshold from below and hence they are lost for stronger $G_C$ attraction.
Note here that we have defined the particle emission threshold as $\mu^*=\min(\abs{\mu},\abs{\mu_C})$ which refers to the minimum excitation energy required to emit a particle from either component.
This crossing entails the self-evaporation process of the three-component mixture. 
Such a finding suggests that the specific collective excitation modes, lying above the particle emission threshold, are not sustained by the three-component droplet configuration. 
A similar behavior but for the collective modes of the 1D single-component droplet has been reported in~\cite{Collective1D}.

Up to this point we studied the properties of three-component particle balanced mixtures, where it was shown that the third one binds to the symmetric two-component  droplet and becomes partially trapped from the latter. 
A question that remains still open is whether the third component can be used as a knob to regulate the droplet structure and if it can be fully trapped by the symmetric  droplet. 
In this latter case, the low density FT tails of component $C$ may vanish or become less extended while the sharp density peaks at the center may become smoother. 
For these reasons, in the following, we explore the impact of particle-imbalance between the symmetric  droplet and component $C$, where a larger atom number occupies the droplet.

\subsection{Particle imbalance: transition from partial to full trapping}\label{ParticleImbalanced}

\begin{figure*}[ht]
\centering   
\includegraphics[width=0.95\linewidth]{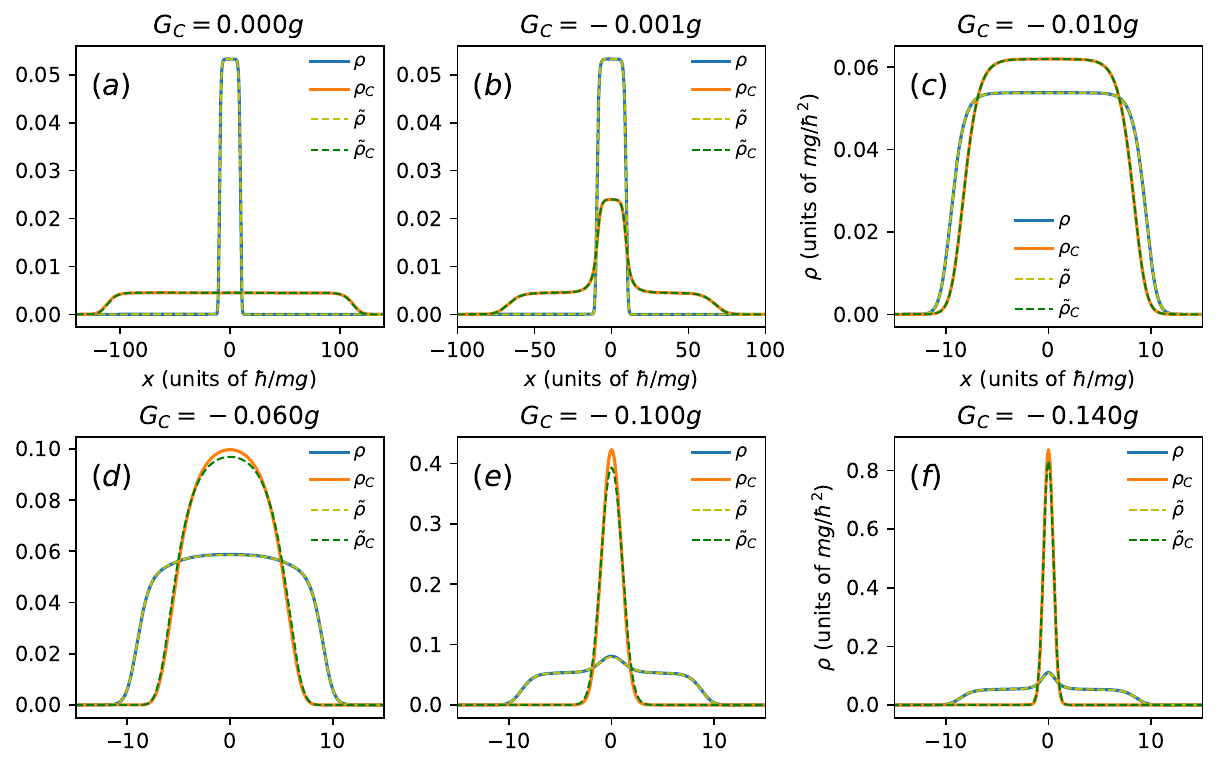}
\caption{Particle imbalanced ground state densities of the majority droplet with $N=10^3$ atoms and the third component containing $N_C=10$ particles, with $g_C=g$, $g_{AB}=-0.96g$, and different intercomponent attractions $G_C$ (see legends).
For vanishing [panel (a)] or very weak [panel (b)] intercomponent attraction $G_C$, the majority droplet is decoupled and loosely bound respectively to the minority component $C$. 
Turning to stronger intercomponent attractions the minority component becomes fully trapped by the majority droplet [panels (c)-(f)], while relatively small undulations build progressively atop the majority droplet density in the neighborhood of component $C$ [panels (e), (f)]. 
Dashed lines represent the solution of the effective model of Eq.~\eqref{eGPEsApprox2}  derived in Sec.~\ref{sec:ParticleImbalance}.}
\label{fig:N1000Densities}
\end{figure*}

It is known that a symmetric droplet  in the absence of a third component (as in our case for $G_C=0$), transitions from a Gaussian to a FT structure for increasing atom number while keeping the involved interactions fixed~\cite{AstrakharchikMalomed1DDynamics,Collective1D}.  
Such a transition behavior occurs also in our setting even in the presence of intercomponent interactions ($G_C=-0.01$) between the droplet and the $C$ component. 
This can be directly inferred from the blue solid lines in Fig.~\ref{fig:ContinuationOverN}(a)-(d) representing the droplet ground state distributions for distinct particle numbers. 
Importantly, as the two majority components ($A$ and $B$) deform toward the FT configuration, they are capable of trapping the minority component $C$, in spite of the small intercomponent attraction $G_C=-0.01g$. 
In particular, the fraction of component $C$ that remains untrapped by the symmetric droplet and occupies a FT 
single-component structure in the particle balanced case [Fig.~\ref{fig:N10Results}(b)] becomes progressively smaller and smoother losing its FT character as $N$ increases [Fig.~\ref{fig:ContinuationOverN}(a), (b)], while simultaneously the droplet acquires a FT configuration.  
Eventually, for sufficiently large particle number the FT droplet of the majority components fully traps the component $C$, see Fig.~\ref{fig:ContinuationOverN}(c), (d). 
We have checked that this transition behavior occurs also for other $G_C$ values, albeit shifted to larger atom numbers of the majority droplet for decreasing magnitude of $G_C$ (i.e., smaller attractions), see also the discussion below and Fig.~\ref{fig:N1000Densities}.

To further understand the aforementioned transition of the three-component mixture we inspect the dependence of the involved chemical potentials which in general take negative values throughout the considered atom number variation as shown in Fig.~\ref{fig:ContinuationOverN}(e). 
Concretely, the chemical potential of the majority (two-component) droplet, $\mu$, decreases sharply for larger particle number and subsequently saturates (for the used parameters around $N\approx 250$) close to its value $\mu^s \approx -1.066$ attained at the decoupled limit ($G_C=0$), see also Sec.~\ref{sec:ParticleImbalance} for further details. 
Beyond this point, increasing the symmetric droplet atom number $N$ further does not affect its chemical potential $\mu$. 
Conversely, the chemical potential $\mu_C$ of the minority component $C$, is initially almost unaffected by the atom number increase of the symmetric droplet being nearly constant at its value at the decoupled limit, i.e., $\mu_C \approx \mu_C^s$. 
This situation takes place as long as component $C$ remains partially trapped by the majority droplet.  
However, a further increase of $N>250$ leads to pronounced and eventually ($N>550$) to full trapping of the minority component $C$ within the symmetric droplet and its (negative) chemical potential features a substantial decrease with $N$, indicating that it progressively becomes more strongly bound. 
This enhancement of the bound state is naturally imprinted in the total chemical potential, $\mu_{\rm {tot}}=\mu+\mu_{C}$, as can be readily seen in Fig.~\ref{fig:ContinuationOverN}(e). 
A complementary discussion of the total energy of the three-component system in terms of the involved interaction parameters is provided in Appendix~\ref{app:phase_diagram}.

\begin{figure}
\centering
\includegraphics[width=0.95\linewidth]{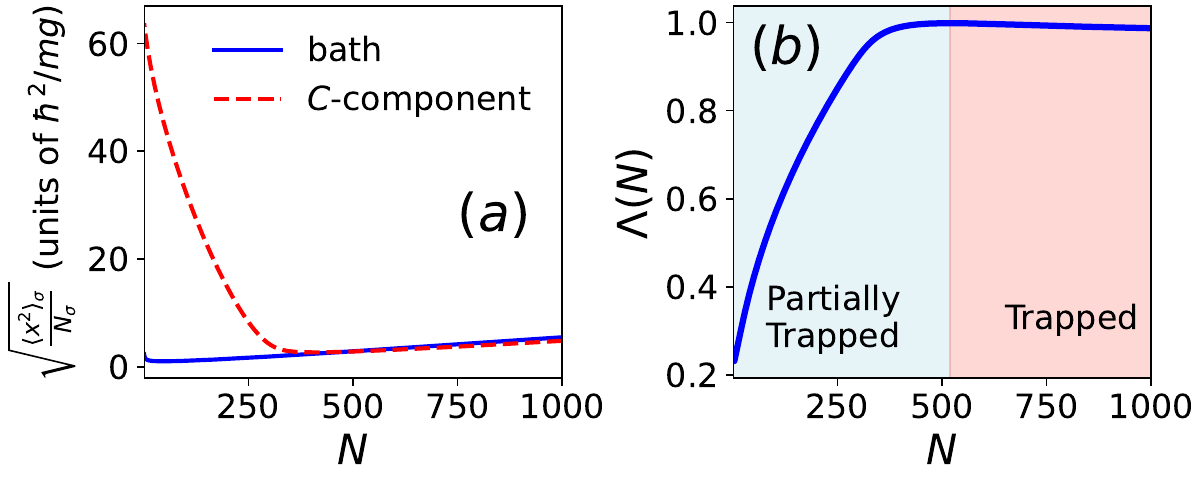}
\caption{(a) Standard deviation of the position of the minority and droplet subsystems (see legend) and (b) the overlap between the droplet and the minority components (see main text) as a function of the droplet atom number. 
The transition from the partially to the fully trapped regimes around $N \approx 500$ is evident in both measures and it is designated by the shaded areas in panel (b).
In all cases, the ground state of the three-component mixture with $N_C=10$, $G_C=-0.01g$, $g=g_C$, and $g_{AB}=-0.96g$ is considered. 
}
\label{fig:trappingVsN}
\end{figure}

We next aim to offer a concrete quantitative characterization of the partial and fully-trapped parametric regimes between the minority component $C$ and the two-component droplet. 
For this study, we employ the normalized standard deviation of the position $\sqrt{\braket{x^2}/N}=\sqrt{\braket{\Psi|x^2|\Psi}/N}$ for the droplet and $\sqrt{\braket{x_C^2}/N_C}=\sqrt{\braket{\Psi|x_C^2|\Psi}/N_C}$ for the minority subsystem.  
This measure, capturing the width of the density distribution of each subsystem, is shown in Fig.~\ref{fig:trappingVsN}(a) with respect to the two-component droplet atom number.  
As discussed above, for smaller particle numbers in the majority components (e.g.  $N<400$), the minority species is not fully trapped by the droplet as it is evident by its much larger standard deviation compared to the droplet, see also Fig.~\ref{fig:ContinuationOverN}(a), (b) for the associated density distributions. 

Indeed, as the minority component becomes progressively more trapped by the droplet, its spatial extent rapidly reduces, until it becomes approximately equal to (e.g. around  $N\approx 500$) or slightly smaller than the droplet size (e.g. at $N>600$), see also Fig.~\ref{fig:ContinuationOverN}(c). 
This indicates that component $C$ resides entirely within the droplet bath for sufficiently large particle number $N$. 
To complement our arguments, we also provide in Fig.~\ref{fig:trappingVsN}(b) the density overlap between the droplet and the minority component $C$, given by $\Lambda=\abs{ \braket{\Psi|\Psi_C} }/\sqrt{NN_C}$. 
Notice that $\Lambda=1$ ($\Lambda=0$) implies full (zero) overlap between the involved components.  
Accordingly, in the partially trapped phase, the overlap rapidly increases as a growing fraction of minority component $C$ becomes trapped by the droplet. 
Around $N\approx 500$ the overlap reaches its maximum value ($\Lambda \approx 1$), as the densities of the minority species and the two-component droplet nearly coincide. 
Finally, upon further increasing the number of atoms in the droplet, the overlap exhibits a slightly decreasing tendency as the minority component localizes further  within the droplet.

Having identified the crossover from partial to full trapping of component $C$ by the majority droplet, it is interesting to explore how this is imprinted in the underlying excitation spectrum and whether it entails modifications in the collective mode behavior. 
The respective numerically computed excitation spectrum in terms of the symmetric droplet atom number using the BdG analysis is illustrated in Fig.~\ref{fig:ContinuationOverN}(f).   
Starting from small particle numbers $N$ of the majority droplet comparable to the minority component $C$, the collective mode frequencies are well separated energetically showing an increasing trend for larger $N$ and cross the particle emission threshold, $\mu^*=\min(\abs{\mu}, \abs{\mu_C})$. 
This crossing takes place at different $N$ values depending on the energy of the collective modes, namely higher-order ones typically cross $\mu^*$ at smaller $N$ compared to energetically lower ones.

Interestingly, there is an atom number parametric window, see $N \in [380,440]$ in Fig.~\ref{fig:ContinuationOverN}(f), where the lowest collective mode lies very close to the particle emission threshold (here $\abs{\omega}/\mu^*\approx 0.993$) and all energetically higher modes have crossed $\mu^*$. 
This suggests that in the crossover region of component $C$ from being partially trapped to fully trapped, the three-component mixture is in a rather delicate balance. 
Namely, a perturbed/excited three-component mixture does not sustain higher-lying collective modes and it rather prefers to emit particles from component $C$ in order to enter the fully trapped regime. 
A further increase of the atom number ($N$) in the majority droplet facilitates the reappearance of excitation modes in the spectrum through crossing $\mu^*$ from above, hence becoming available again to the three-component mixture. 
Importantly, the behavior of these modes is somewhat different compared to the ones that were present below the crossover region.  
Indeed, below the crossover region (e.g., $N<300$), the modes are gapped (aside from their sign degeneracy), while (at least) the first few lower lying ones above this region (e.g. $N>500$) feature a pairing tendency becoming two-fold quasi-degenerate (on top of their trivial sign degeneracy). 
It is also worth mentioning that our simulations delineate that the above-discussed transition region is rather smooth and it is not easy to define a specific transition point.

\begin{figure}
\centering
\includegraphics[width=0.99\linewidth]{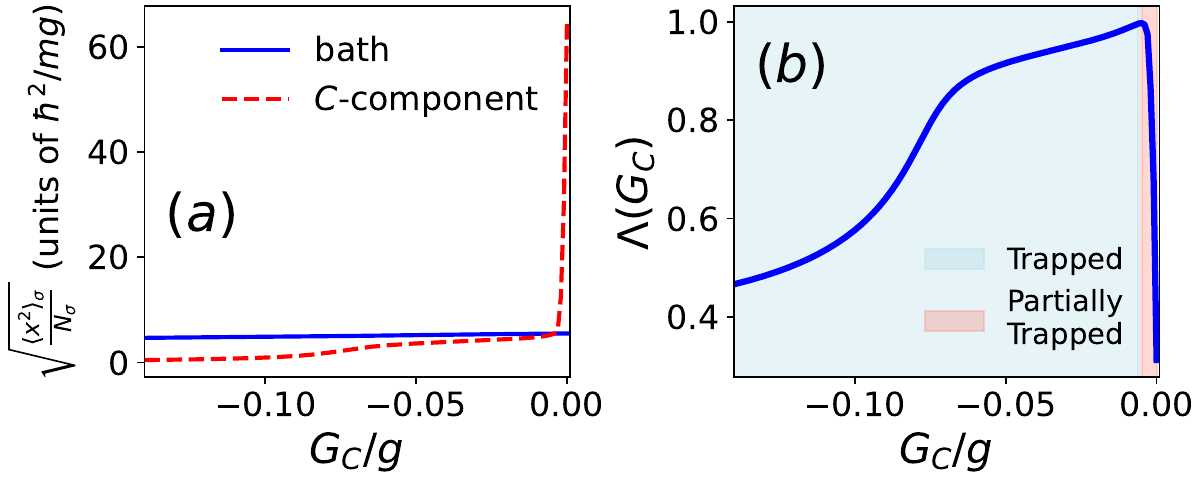}
\caption{Ground state (a) standard deviation of the position of the two-component droplet and the minority species (see legend) and (b) spatial overlap between the droplet and component $C$ for varying intercomponent coupling $G_C$ across the attractive MF stability regime. 
The transition from the partially to fully and afterwards to the partially trapped regimes is imprinted in both measures and it is marked by the shaded areas in panel (b).  
The three-component mixture contains $N_C=10$, $N=1000$ bosons with $g_C=g$, $g_{AB}=-0.96g$.
}
\label{fig:trappingVsG}
\end{figure}

\begin{figure*}[ht]
\centering   
\includegraphics[width=0.95\linewidth]{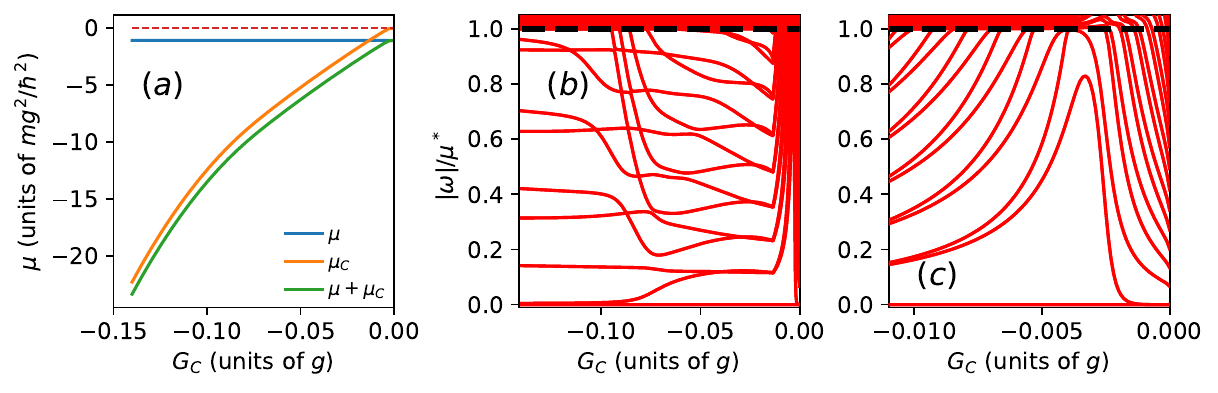}
\caption{(a) Chemical potentials (see legend) and (b) excitation spectrum of a particle imbalance three-component mixture as a function of the intercomponent attraction between the majority droplet with $N=10^3$ and component $C$ having $N_C=10$ atoms.   
It is observed that upon increasing $G_C$, component $C$ becomes less strongly bound as indicated by the decreasing magnitude of $\mu_C$, while the chemical potential of the majority droplet ($\mu$) remains largely unaffected. 
The horizontal red dashed line represents $\mu=0$. 
In the spectrum, it is found that the collective modes show a rather complicated interaction-dependent behavior in terms of $G_C$ [panel (b)] experiencing multiple avoided crossings.   
(c) Same as (b), but restricted to weak intercomponent attractions to clearly visualize the behavior of the collective modes in this interaction regime. 
As in the case of varying $N$ depicted in Fig.~\ref{fig:ContinuationOverN}(f), the modes (except the lower-lying one) cross the particle emission threshold, $\mu^*=\min(\abs{\mu}, \abs{\mu_C})$ (see dashed horizontal line), from below and afterwards they re-emerge from above.  
Other interaction parameters correspond to $g_C=g$, and $g_{AB}=-0.96g$.}
\label{fig:N1000Energies}
\end{figure*}

Up to this point, we studied the phases of particle-imbalanced three-component mixtures for varying atom numbers in the symmetric droplet and relatively weak intercomponent attraction ($G_C$) between the majority droplet and the minority species. 
In what follows, we examine the impact of $G_C$ on the ensuing ground-state configurations for fixed particle imbalance by setting $N=10^3$ and $N_C=10$, see Fig.~\ref{fig:N1000Densities}. 
The ensuing phases are associated with the three-component setting and do not bear two-component analogues. 
We observe that for vanishing or sufficiently small intercomponent coupling e.g., $G_C = 0$ [Fig.~\ref{fig:N1000Densities}(a)] or $G_C = -0.001g$ [Fig.~\ref{fig:N1000Densities}(b)], the distribution of component $C$ is practically decoupled from the symmetric droplet being completely untrapped and loosely trapped by the symmetric droplet respectively.  
Namely, if $G_C=0$ component $C$ configures in a FT distribution being fully decoupled from the FT configuration of the majority droplet, see also the discussion in Sec.~\ref{sec:ParticleImbalance}. 
Turning to relatively small $G_C<0$ values, the configurations of both the droplet and component $C$ change noticeably. 
The symmetric droplet maintains its FT shape and its width becomes slightly smaller, while the atom cloud of component $C$ shrinks significantly accommodating a fraction of atoms being trapped in the majority droplet and the remaining atoms residing at the delocalized density tails similarly to the particle-balanced case [Fig.~\ref{fig:N10Results}(b)].

For larger attractions, such as $G_C=-0.01g$ or $G_C=-0.06g$ illustrated in Fig.~\ref{fig:N1000Densities}(c), (d), the density undulations of component $C$ disappear with the latter becoming fully trapped within the majority droplet.  
The density of the minority component $C$ becomes more spatially localized transitioning from a FT for $G_C=-0.01g$ [Fig.~\ref{fig:N1000Densities}(c)] to a Gaussian shape for $G_C=-0.06g$ [Fig.~\ref{fig:N1000Densities}(d)). Accordingly, the majority droplet density slightly shrinks due to the increased attraction but it maintains a FT configuration enclosing the minority component. 
A further increase of the intercomponent attraction, e.g., to $G_C=-0.1g$ or $G_C=-0.14g$ as depicted in Fig.~\ref{fig:N1000Densities}(e), (f) yields dramatic changes in the densities of both the symmetric  droplet and component $C$. 
We remark that these values of $G_C$ refer to strong attractions when compared to the MF stability edge $\abs{G_C} \leq 0.1 \sqrt{2}$. Specifically, the minority component features a substantial localization arranging in a sharply peaked soliton-like structure, while being fully trapped in the symmetric droplet which forms a comparatively spatially extended FT distribution. 
The latter exhibits noticeable density modulation at the vicinity of the minority cloud due to the appreciable attraction, similarly to two-component particle imbalanced droplets discussed in Ref.~\cite{Englezos2024}.

It is useful to  explicitly visualize the transition from partially to fully trapped regimes of the minority species within the two-component droplet. To this end, we present the standard deviation of the position of each of these subsystems in Fig.~\ref{fig:trappingVsG}(a) as a function of the intercomponent coupling $G_C$ throughout the attractive MF stability regime. 
For sufficiently weak attractions ($-0.005<G_C<0$) the minority species is partially untrapped exhibiting a much larger spatial extent as compared to the droplet, see also Fig.~\ref{fig:N1000Densities}(a).
However, increasing the magnitude of the attraction, e.g., in the interval $-0.08<G_C<-0.005$ favors spatial localization of the minority species inside the droplet, see also Fig.~\ref{fig:N1000Densities}(b), (c). 
This behavior can be readily inferred by the associated sharply decreasing standard deviation of the position of component $C$, which quickly becomes smaller than the one corresponding to the droplet bath. 
Further increasing the attraction, $G_C<-0.08$, results in a sigmoid-like decrease of the width of the minority species which practically transitions from a Thomas-Fermi like density profile [Fig.~\ref{fig:N1000Densities}(d)] toward a soliton-like one [Fig.~\ref{fig:N1000Densities}(e), (f)]. 
The transition from partially to fully trapped regimes for increasing attraction is also deduced from the behavior of the spatial overlap, $\Lambda(G_C)$, between component $C$ and the droplet shown in Fig.~\ref{fig:trappingVsG}(b).  
Specifically, $\Lambda(G_C)$ exhibits a sharp increase for weak attraction where the minority component progressively localizes and becomes fully trapped by the droplet. 
At the transition region towards the fully trapped regime ($G_C\approx-0.005$), the densities of the droplet and the minority species nearly coincide and $\Lambda(G_C) \approx 1$. 
Finally, further increasing the intercomponent attraction $G_C$ leads to a sigmoid decreasing trend of $\Lambda(G_C)$ in line with the behavior of the standard deviation of the position, discussed above [Fig.~\ref{fig:trappingVsG}(a)].

The degree of the bound state character of the aforementioned three-component states along with the binding of the constituting components can be further estimated through the individual chemical potentials presented in Fig.~\ref{fig:N1000Energies}(a) for varying $G_C$.  
As expected, the chemical potential of the majority (two-component) droplet $\mu$ is negative for every $G_C$ since the droplet is a self-bound configuration. 
Interestingly, $\mu$ remains approximately constant at its saturation value $\mu^s\approx -1.066$ at the decoupled ($G_C=0$) limit and it appears to be largely insensitive to the intercomponent coupling with the minority component $C$. 
We may attribute this insensitivity to the large particle imbalance. 
Recall here that in the particle-balanced scenario illustrated in Fig.~\ref{fig:N10Results}(c) the situation is reversed and $\mu$ increases with increasing $G_C$.  
Simultaneously, the chemical potential of the minority component ($\mu_C$) has a more involved behavior. 
Indeed, at $G_C=0$ it is slightly negative acquiring its saturation value $\mu_C^s \approx -0.0225$, and as long as $G_C \neq 0$ it monotonically decreases with decreasing $G_C$.
This implies the formation of stronger bound configurations of component $C$, while the crossing point between $\mu$ and $\mu_C$ at $G_C\approx -0.0135g$ happens when the transition from partially to fully trapped component $C$ occurs. 
We remark that the total chemical potential $\mu_{\rm {tot}}=\mu+\mu_C$  mainly inherits the behavior of $\mu_C$ suggesting stronger binding of the three-component setting for larger attractions.

Finally, for completeness we examine the excitation spectrum of the system as a function of $G_C$ whose real part is depicted in Fig.~\ref{fig:N1000Energies}(b), (c) aiming to identify distinct spectral signatures for the crossover from partially to fully trapped component $C$ within the symmetric droplet. 
As far as the configurations discussed in the previous sections are concerned, the imaginary parts of the corresponding eigenvalues are zero suggesting that the underlying configurations are dynamically stable. 
It can be readily seen in Fig.~\ref{fig:N1000Energies}(b) that the collective mode branches have a strong interaction dependence which leads to a rather complicated behavior. 
In particular, for strong intercomponent attraction $G_C \lessapprox -0.1g$, a sequence of avoided crossings between consecutive modes is observed. 
Decreasing the intercomponent attraction towards $G_C \lessapprox -0.05g$ results in several additional modes crossing from above the particle emission threshold and featuring avoided crossings with the pre-existent lower-lying modes along with a pairing tendency [Fig.~\ref{fig:N1000Energies}(b)]. 
Turning to sufficiently weak attractions, such as $G_C \gtrapprox -0.02g$, energetically neighboring collective modes come very close and subsequently cross the particle emission threshold [Fig.~\ref{fig:N1000Energies}(c)] with only the lower-lying one remaining below that threshold. 
In this regime the transition from fully to partially trapped component $C$ by the symmetric droplet occurs. It is interesting to note that the aforementioned mode-couplings through the avoided crossings are inherently related to the three-component mixture and the presence of all LHY contributions used. 
Namely, these crossings are absent either if the LHY term related to the third component is not taken into account (see also Appendix~\ref{app:MF}) or in two-component droplet settings with all LHY proper contributions~\cite{charalampidis2024two}. 
The former observation implies the crucial stabilization role of the LHY term of the third component which is currently elusive in the literature. 
The second point suggests that the impurity through its coupling to the two-component droplet can act as a ``sensor" for probing the collective excitations of the latter, a mechanism that should be further investigated in forthcoming 
studies.

Next, as $G_C \to 0$, the higher-lying modes re-emerge, crossing the particle emission threshold from above and forming unpaired gapped energy branches as component $C$ starts to decouple from the majority (two-component) droplet [Fig.~\ref{fig:N1000Energies}(c)]. 
This behavior is consistent with that illustrated in Fig.~\ref{fig:ContinuationOverN}(f), where increasing the atom number at fixed $G_C$  resulted in the opposite transition, as the minority component $C$ became trapped by the majority droplet. 
Overall, the complex interaction-dependent behavior of the ensuing collective modes suggests that the particle-imbalanced three-component mixture may host rich dynamical phenomena, e.g. upon quenching or sweeping the intercomponent coupling strength. 
This is certainly an intriguing direction for future studies where the nonequilibrium features imprinted on the density profiles of the individual components can be associated e.g. with the observed avoided crossings in the spectrum.

\section{Effective extended Gross-Pitaevskii Equation models at high particle imbalance}\label{sec:ParticleImbalance}

To gain further insights on the character of the three-component states discussed in the previous sections, we next develop relevant effective models based on suitable reductions of the full eGPEs~(\ref{eGPETwoPlusOneComponent}). 
More concretely, we consider the case where the third component ($C$) is occupied by a much smaller number of atoms compared to the two symmetric  components $A$ and $B$ forming the majority droplet, i.e. $N \gg N_C \gg 1$.
Additionally, in this context, all three components remain macroscopically occupied, such that the Bogoliubov approximation remains justified. 
Taylor expanding with respect to $N_C/N$ the elements in the LHY energy density described by Eq.~\eqref{ELHY2Plus1} (more formally called a Puiseux expansion~\cite{willis2008compute}), and keeping only terms scaling with non-negative powers of $N$ yields the approximate expression 
\begin{equation}
\label{ApproxE_LHY}
\begin{split}
\mathcal{E}_{\rm LHY}^\prime  =& \frac{-2\sqrt{mg^3}}{3\hbar\pi} \Bigg[ 
(f_{-}^{\frac{3}{2}} + f_{+}^{\frac{3}{2}}) n^{\frac{3}{2}} + \frac{3f_G}{4}  \sqrt{f_{+}n}n_C \\  
&
+\big( \frac{g_C}{g} - \frac{f_G}{2} \big)^{\frac{3}{2}} n_C^{\frac{3}{2}} + \mathcal{O}(\frac{1}{\sqrt{N}}) \Bigg].      
\end{split}    
\end{equation}
Here, the density independent parameters $f_\pm = 1\pm g_{AB}/g$ and $f_G= 4 G_C^2/ (g^2f_{+})$, see also Appendix~\ref{app:limit} for further details on this derivation. 
It is also instructive to remark that following this notation, the MF stability window translates to $0<f_{\pm}<2$ and $f_G\leq2g_C/g$. 

The scaling of the individual terms participating in the total approximate energy density, i.e., 
\begin{equation}
\label{ApproxEnergy}
\mathcal{E}^\prime = gn^2 + \frac{g_C}{2} n_C^2 + 2G_C nn_C + g_{AB}n^2 + \mathcal{E}_{LHY}^\prime,
\end{equation}
and the corresponding eGPEs with the particle number is summarized in Table~\ref{tab:scalingExample} for an imbalanced setting hosting $N=10^3$ and $N_C=10$ atoms.
\begin{table}[h]
\centering
\begin{tabular}{|c|c|c|c|c|c|c|c|c|}
\hline $\mathcal{E}\sim $ & $N^2$ & $N N_C$ & $N^{\frac{3}{2}}$ & $\sqrt{N}N_C$ & $N_C^2$ &  $N_C^{\frac{3}{2}}$ & $\frac{N_C^{\frac{5}{2}}}{N}$ & $\frac{N_C}{\sqrt{N}}$\\ \hline 
Energy & $10^6$ & $10^4$ & $10^{\frac{9}{2}}$ & $ 10^{\frac{5}{2}}$ & $10^2$ &  $10^{\frac{3}{2}}$ & $\frac{1}{10^{\frac{1}{2}}}$ & $\frac{1}{10^{\frac{1}{2}}}$ \\ \hline
$\frac{\partial\mathcal{E}}{\partial n}\sim $ & $N$ & $N_C$ & $N^{\frac{1}{2}}$ & $\frac{N_C}{\sqrt{N}}$ & $0$ &  $0$ & $\frac{N_C^{\frac{5}{2}}}{N^2}$ & $\frac{N_C}{N^{\frac{3}{2}}}$\\ \hline
$\rm eGPE_{\rm (A,B)}$  & $10^3$ & $10$ & $10^{\frac{3}{2}}$ & $ \frac{1}{10^\frac{1}{2}}$ & $0$ &  $0$ & $\frac{1}{10^\frac{5}{2}}$ &$\frac{1}{10^\frac{5}{2}}$ \\ \hline
$\frac{\partial\mathcal{E}}{\partial n_C}\sim $ & $0$ & $N$ & $0$ & $\sqrt{N}$ & $N_C$ &  $\sqrt{N_C}$ & $\frac{N_C^{\frac{3}{2}}}{N}$ & $\frac{1}{\sqrt{N}}$\\ \hline
$\rm  eGPE_{\rm (C)}$ & $0$ & $10^3$ & $0$ & $ 10^{1.5}$ & $10$ &  $10^{\frac{1}{2}}$ & $\frac{1}{10^\frac{3}{2}}$ &$\frac{1}{10^\frac{3}{2}}$ \\ \hline
\end{tabular}
\caption{Scaling of the various terms appearing in the total approximate energy density [Eq.~(\ref{ApproxEnergy})] containing the LHY contribution given by Eq.~(\ref{ApproxE_LHY}) and the corresponding eGPEs [Eq.~(\ref{eGPETwoPlusOneComponent})] with respect to the atom number for a highly particle imbalanced three-component mixture containing $N=10^3$, and $N_C=10$. It is evident that terms encompassing the majority droplet atom number, $N$, have the dominant contribution.}
\label{tab:scalingExample}
\end{table}
As can be readily seen, in all cases the dominant terms are those involving the majority components ($A$ and $B$) atom number, while the ones associated with the minority component particle number are significantly smaller.
Finally, the last two columns in Table~\ref{tab:scalingExample} have an inverse scaling with $N$ and they represent the first neglected terms in the expansion of the LHY energy of Eq.~\eqref{ApproxE_LHY}.
Clearly, their contribution is indeed negligible compared e.g. to the MF terms provided in the first, second, and fifth columns of Table~\ref{tab:scalingExample}. 
Hence, these LHY contributions can be safely omitted.

\begin{figure}[ht]
\centering   \includegraphics[width=0.9\linewidth]{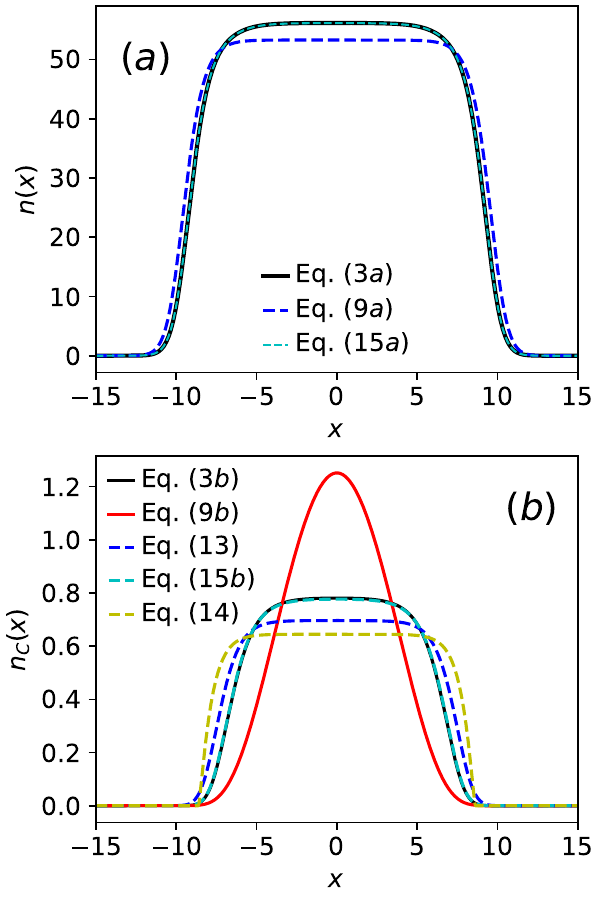}
\caption{Comparison between the predictions of different reduced eGPE models introduced in Sec.~\ref{sec:ParticleImbalance} (see legends) and the full numerical solution of the eGPEs of Eq.~(\ref{eGPETwoPlusOneComponent}). 
Ground-state density of (a) the majority droplet and (b) the minority $C$ component. 
Clearly the predictions of the coupled model of Eq.~\eqref{eGPEsApprox2} are in excellent agreement with the exact solutions. 
Neglecting the MF coupling results in noticeable deviations, however the model of an interacting minority component (Eq.~\eqref{effectiveEGPEMinority}) dressed by and immersed in the symmetric two-component droplet (Eq.~\eqref{eGPEsApprox1a}) is fairly close to the exact results. 
A highly particle imbalanced three-component system is considered with  $N=10^3$, $N_C=10$ atoms, as well as $g=g_C=g$, $g_{AB}=-0.96g$, and $G_C=-0.04g$ interaction strengths. 
}
\label{fig:Reduction}
\end{figure}

\subsection{Decoupled limit}\label{sec:Decoupled}

As a first, somewhat crude approximation, we assume that the particle imbalance is sufficiently high such that the majority droplet component is practically unaffected by the presence of the minority species. 
As such, it is possible to neglect all terms in the LHY energy density of Eq.~(\ref{ApproxE_LHY}) that scale with $N_C$ (and hence also intracomponent interactions of component $C$) and arrive in a reduced $\mathcal{E}^\prime_{{\rm red}}$ out of the approximate $\mathcal{E}'$ given by Eq.~\eqref{ApproxEnergy}.
Following a standard variational treatment for the energy density $\mathcal{E}^\prime_{{\rm red}}$ results in the approximate eGPEs 
\begin{subequations}
\label{eGPEsApprox1}
\begin{align}
\label{eGPEsApprox1a}
i\hbar\frac{\partial \Psi^\prime}{\partial t}& = - \frac{\hbar^2}{2m} \frac{\partial^2 \Psi^\prime}{\partial x^2} +\Big( (g+g_{AB}) \abs{\Psi^\prime}^2\\
&-\frac{\sqrt{m}}{2\hbar\pi}  
(f_{-}^{3/2} + f_{+}^{3/2}) g^{3/2}\abs{\Psi^\prime} \Big)\Psi^\prime, \nonumber  \\ \label{eGPEsApprox1b}
i\hbar\frac{\partial \Psi^\prime_C}{\partial t}& = \Big(- \frac{\hbar^2}{2m} \frac{\partial^2}{\partial x^2} + V_{{\rm eff}}(n^\prime) \Big)\Psi^\prime_C, 
\end{align}
\end{subequations}
where $n^\prime(x) = \abs{\Psi^\prime(x)}$ and $n_C^\prime(x) = \abs{\Psi_C^\prime(x)}$.
These expressions reveal that in this limit and to lowest-order, the majority droplet satisfies a single-component eGPE, while the minority component is under the influence of an effective box-like potential 
\begin{equation}
V_{{\rm eff}}(n^\prime) = 2G_{C} n^\prime - \frac{\sqrt{f_{+}mg^3}}{2\hbar\pi}f_G  \sqrt{n^\prime},
\label{Veff}  
\end{equation}
similarly to the two-component particle imbalanced droplet mixture discussed in Ref.~\cite{Englezos2024}. 
Apparently, this is created by the majority droplet and depends on the intercomponent interaction ($G_C$) between the droplet and component $C$ but also on the interaction parameters ($g$, $g_{AB}$) of the droplet. 
The majority droplet, therefore, acts as an effective attractive potential for component $C$, whose depth is tuned by $G_C$ while its shape is inherited from the majority density. Increasing $\abs{G_C}$ consequently strengthens the confinement experienced by the minority species,  promoting its spatial localization within the (two-component) droplet. 
Recall that similar effective potentials in highly particle  imbalanced mixtures have been successfully used in completely different contexts such as polarons~\cite{PolaronDynamics2}, rogue waves~\cite{romero2024experimental,bougas2025observation}, and the Townes soliton~\cite{BakkaliTowns2021,bakkali2022townes}.

This simplified approach motivates us to search for stationary solutions, by setting $\Psi^\prime = e^{-i\mu^\prime t/\hbar}\Phi^\prime$ and $\Psi^\prime_C = e^{-i\mu_C^\prime t/\hbar}\Phi^\prime_C$.
Accordingly, Eq.~\eqref{eGPEsApprox1a} describing the majority droplet admits the standard droplet solution~\cite{Petrov2015,PetrovLowD} being valid in the thermodynamic limit 
\begin{equation}\label{DropletSolution}
\Phi^\prime(x) = \frac{A}{1+\Gamma \cosh{\lambda x}},
\end{equation}
where the involved parameters correspond to 
\begin{subequations}
\label{MajorityDropletSolution}
\begin{align}
A =& \frac{-6\hbar\pi \mu^\prime}{\sqrt{mg^3}(f_{-}^{3/2} + f_{+}^{3/2} )} = \frac{\sqrt{n^s}\mu^\prime}{{\mu^{s}}}, \\
\lambda =& \sqrt{\frac{-2m\mu^\prime}{\hbar^2}}, \\
\Gamma =& \sqrt{1 + \frac{36\hbar^2\pi^2 \mu^\prime}{mg^3(f_{-}^{3/2} + f_{+}^{3/2} )^2}} = \sqrt{1-\frac{\mu^\prime}{\mu^{s}}}.
\end{align}
\end{subequations}
In these expressions, the droplet saturation density is given by $n^s = \frac{mg^3 (f_{-}^{3/2} + f_{+}^{3/2} )^2 }{9\hbar^2\pi^2(g+g_{AB})^2}$, with associated saturation value of the chemical potential $\mu^s = -n^s(g+g_{AB})/2$. 
Note that $n^s$ and $\mu^s$ appearing in Eq.~\eqref{MajorityDropletSolution} coincide with the values at the decoupled limit $G_C$. 
This solution is valid within the interval $\mu^s <  \mu^\prime < 0$~\cite{PetrovLowD,Englezos2025Droplets}. 
For sufficiently large particle imbalance and relatively weak coupling $G_C$, this solution adequately captures the exact ground-state obtained by numerically solving the exact eGPEs of Eq.~\eqref{eGPETwoPlusOneComponent}, as shown in Fig.~\ref{fig:Reduction}(a).
The observed discrepancies in the peak amplitude and width of the droplet distribution are traced back to the neglected MF intercomponent coupling ($G_C$) with the minority component. 
Indeed, as it will be showcased below, by introducing the MF coupling in Eq.~(\ref{eGPEsApprox1a}) again, it leads to an excellent agreement with the full eGPEs predictions. 
We remark, on the other hand, that for larger atom numbers in the droplet e.g. $N=10^4$ the deviations reduce (not shown) even in the absence of $G_C$. 
This is a consequence of the fact that the terms $\sim \abs{\Psi^\prime}^2$ and $\sim \abs{\Psi^\prime}$ in Eq.~(\ref{eGPEsApprox1a}) dominate.

Substituting the exact droplet solution given by Eq.~\eqref{DropletSolution} (along with Eq.~\eqref{MajorityDropletSolution}) into Eq.~\eqref{eGPEsApprox1b} describing the minority component, we can readily diagonalize the linear system $H \Phi^{^\prime C}_n = - \frac{\hbar^2}{2m} \frac{\partial^2 \Phi^{^\prime C}_n}{\partial x^2} + V_{{\rm eff}}(n)\Phi^{^\prime C}_n = E_n^C \Phi^{^\prime C}_n, $ numerically, and find its eigenstates. 
The latter are similar to the ones of a single-particle trapped in a finite box potential with depth dictated by $V_{{\rm eff}}(n^s)$, due to the FT shape of the majority droplet density. 
Hence, they prominently deviate from the full eGPE predictions for the shape of the minority component, see Fig.~\ref{fig:Reduction}(b). 
Thus, it becomes clear that this level of approximation is too restrictive for approaching the distribution of the minority component. 
Indeed, the MF and the LHY nonlinear terms, albeit having a smaller magnitude compared to $V_{{\rm eff}}(n^s)$, are responsible for giving rise to qualitatively different structures. 
As such, in order to improve the accuracy of our model, even in the decoupled limit, we take into account the MF intracomponent repulsion ($g_C\abs{\Psi_C}^2\Psi_C$) and the LHY term $\sim \abs{\Psi_C}\Psi_C$ in Eq.~\eqref{eGPEsApprox1b}. 
This leads to the effective single-component eGPE governing the minority component 
\begin{equation}
\label{effectiveEGPEMinority}
\begin{split}
i\hbar\frac{\partial \Psi^{\prime\prime}_C}{\partial t}& = - \frac{\hbar^2}{2m} \frac{\partial^2 \Psi^{\prime\prime}_C}{\partial x^2} + V_{{\rm eff}}(n^\prime)\Psi^{\prime\prime}_C + g_C\abs{\Psi^{\prime\prime}_C}^2\Psi^{\prime\prime}_C \\
&-\frac{\sqrt{mg^3}}{\hbar \pi} \Big( \frac{g_C}{g} - \frac{f_G}{2} \Big)^{3/2} \abs{\Psi^{\prime\prime}_C}\Psi^{\prime\prime}_C,
\end{split}
\end{equation}
where, $\rho^{\prime}(x)=n^{\prime}(x)/N=\Psi^{\prime\prime}(x)/N$ and $\rho^{\prime\prime}_C(x)=n^{\prime\prime}_C(x)/N=\Psi^{\prime\prime}_C(x)/N_C$.

It is important to notice that the symmetric droplet is influencing the minority component in three distinct ways.
Indeed, within the effective potential $V_{{\rm eff}}(n^\prime)$ there is: i) the direct MF intercomponent coupling ($2G_C \abs{\Psi^\prime}^2$), and ii) the beyond MF intercomponent coupling ($\sim \abs{\Psi^\prime}$) stemming from the genuinely three-component LHY energy Eq.~(\ref{ApproxEnergy}). 
Moreover, iii) the beyond MF intracomponent coupling of the minority species is modified by the presence of the majority droplet resulting in a different multiplicative factor $\sqrt{mg^3}(g_C/g- f_G/2)^{3/2}/(\hbar \pi)$ compared to the genuinely single-component LHY factor [$\sim \sqrt{mg_C^3}/(\pi\hbar)$].
The numerical solution of Eq.~\eqref{effectiveEGPEMinority} is shown in Fig.~\ref{fig:Reduction}(b) against the outcome of the full eGPEs of Eq.~\eqref{eGPETwoPlusOneComponent}. 
It qualitatively captures the distribution of the minority component but with a smaller peak density and slightly larger width. 
Despite these deviations the prediction of this model is clearly improved compared to the previous case where the nonlinear terms are absent, compare in particular red solid and blue dashed lines against the black solid one in Fig.~\ref{fig:Reduction}(b).

To obtain analytical insights on the behavior of the minority component, we next consider the Thomas-Fermi limit reached by  neglecting the kinetic term in Eq.~\eqref{effectiveEGPEMinority}. This solution takes the form 
\begin{equation}
\begin{split}
\label{TFlimit}
\Phi_C^{\rm TF} = &\frac{\sqrt{2mg^3}}{8\hbar\pi g_C}(2\frac{g_C}{g} - f_G )^{3/2} \Big[ 1 -U \Big],
\end{split}
\end{equation}
where $U =  (1 + 32\frac{\hbar^2 \pi^2 g_C (\mu_C - V_{{\rm eff}})}{m g^3(2\frac{g_C}{g} - f_G )^{3} } )^{\frac{1}{2}}$.
Following the standard calculation in the Thomas-Fermi limit, we can find the associated density boundary (dubbed here $x_b$) by demanding $\Phi_C^{\rm TF}= 0 \xrightarrow{} \mu^{\rm TF}_C=V_{{\rm eff}}(x_{b})$, while the density is taken to be zero outside this  boundary.
The chemical potential $\mu_C$ is fixed by the normalization condition $N_C=\int_{-x_{b}}^{x_{b}} \abs{\Phi_C^{\rm TF}}^2 dx$.
The boundary is given by $x_{b} = \frac{1}{\lambda}   
\cosh^{-1}{\frac{A-\abs{K}}{\Gamma \abs{K}}}$, where $\abs{K} = \frac{\sqrt{f_{+}mg^3}f_G}{8\hbar\pi G_C} \Bigg[ 1 - \sqrt{1 + \frac{32 G_C \mu^{\rm TF}_C \hbar^2 \pi^2}{f_{+} m g^3 f_G^2 }} \Bigg]$ and $A,\; \lambda,\; \Gamma$ are the same as in Eq.~\eqref{MajorityDropletSolution}.
The maximum density of the minority component is given by substituting $V_{{\rm eff}}=V_{{\rm eff}}(x=0)=V_{{\rm eff}}(n^s)$.
Note that $\Phi_C^{\rm TF}$ is negative and real for $\mu^{\rm TF}_C > V_{{\rm eff}}$, while its real part becomes positive and it obtains a negative imaginary part as well for $\mu^{\rm TF}_C < V_{{\rm eff}}$, i.e. outside the density boundaries $\abs{x}>x_{b}$. 
The Thomas-Fermi profile of Eq.~(\ref{TFlimit}) has a qualitatively similar shape to the exact density [Eq.~\ref{eGPETwoPlusOneComponent}] of the minority component but exhibits a reduced (larger) peak amplitude (width), see the dashed yellow line in Fig.~\ref{fig:Reduction}(b). 
Notice that its deviation from the exact eGPE result is somewhat larger compared to the one of the model of  Eq.~(\ref{effectiveEGPEMinority}) which in addition incorporates the kinetic term. 
This implies that the kinetic energy contribution is important for describing the minority component.

\subsection{Reintroducing mean-field coupling}

The term that we have neglected so far in the approximate eGPE for the two-component droplet of Eq.~\eqref{eGPEsApprox1a} which does not scale with negative powers of $N$ but with $N_C$ is the direct intercomponent MF coupling $\propto G_C n_C$. 
To incorporate this term we employ the variational treatment of the energy density $\mathcal{E}'$ resulting in the following coupled set of approximate eGPEs 
\begin{subequations}
\label{eGPEsApprox2}
\begin{align}
\label{eGPEsApprox2a}
i\hbar\frac{\partial \tilde{\Psi}}{\partial t} =& - \frac{\hbar^2}{2m} \frac{\partial^2 \tilde{\Psi}}{\partial x^2} +\Big( (g+g_{AB}) \abs{\tilde{\Psi}}^2 +  G_C \abs{\tilde{\Psi}_C}^2  \nonumber \\
&-\frac{\sqrt{m}}{2\hbar\pi}  (f_{-}^{3/2} + f_{+}^{3/2}) g^{3/2}\abs{\tilde{\Psi}} \Big)\tilde{\Psi},  \\ \label{eGPEsApprox2b}
i\hbar\frac{\partial \tilde{\Psi}_C}{\partial t} =& - \frac{\hbar^2}{2m} \frac{\partial^2 \tilde{\Psi}_C}{\partial x^2} + g_C\abs{\tilde{\Psi}_C}^2 \tilde{\Psi}_C \nonumber \\
&+ \Big( 2G_{C}\abs{\tilde{\Psi}}^2 - \frac{\sqrt{f_{+}mg^3}}{2\hbar\pi} f_G \abs{\tilde{\Psi}} \Big)\tilde{\Psi}_C \\
&-\frac{\sqrt{2mg^3}}{4\hbar \pi} \Big( 2 \frac{g_C}{g} -f_G \Big)^{3/2} \abs{\tilde{\Psi}_C}\tilde{\Psi}_C . \nonumber
\end{align}
\end{subequations}
In these expressions, $\tilde{\rho}(x)=\tilde{n}(x)/N=\abs{\tilde{\Psi}(x)}^2/N$ and $\tilde{\rho}_C(x)=\tilde{n}_C(x)/N=\abs{\tilde{\Psi}_C(x)}^2/N_C$.
Note here that Eq.~\eqref{eGPEsApprox2b} is identical to that of Eq.~\eqref{effectiveEGPEMinority}, while Eq.~\eqref{eGPEsApprox2a} differs from Eq.~\eqref{eGPEsApprox1a} by a single term, namely $G_C|\Tilde{\Psi}_C|^2$.
This difference, however, results in a coupled set of equations, meaning that the majority droplet density cannot be considered as a static effective potential, and the two equations need to be solved concurrently and self-consistently.
In this sense, the intercomponent MF coupling, introduces the back-action of the minority component onto the majority droplet. 

The numerical solution of these coupled approximate eGPEs is found to be in nearly perfect agreement with the solutions of the exact eGPEs described by Eq.~\eqref{eGPETwoPlusOneComponent}, for weak intercomponent attractions ($G_C<0$), compare the dashed and solid lines in Fig.~\ref{fig:Reduction} and in Fig.~\ref{fig:N1000Densities}(a)-(c). 
This confirms the mutual feedback between the components. Indeed, while the majority droplet provides the effective trapping potential experienced by the  localized component $C$, the backaction of the latter to the majority species through $G_C$  leads to a local deformation of their density profile. 
Note that in the decoupled limit $G_C=0$ [Fig.~\ref{fig:N1000Densities}(f)], the standard droplet solution  [Eq.~\eqref{DropletSolution}] describes the majority droplet and the minority component $C$ with the appropriate parameter (e.g., $\mu \to \mu_C$, $\lambda \to \lambda_C$) modifications for the latter case. 
For larger attractions, as the ones depicted in Fig.~\ref{fig:N1000Densities}(d)-(f), we know that within the full eGPEs [Eqs.~(\ref{eGPETwoPlusOneComponent})], structural deformations of the majority droplet density take place due to the presence of component $C$, yielding a locally modulated FT profile of the majority droplet in the vicinity of component $C$ which has a Gaussian shape. 
While these distorted states cannot be captured by the standard droplet solution of Eq.~\eqref{DropletSolution}, they can be fairly well reproduced by the coupled set of approximate eGPEs [Eq.~\eqref{eGPEsApprox2}], as can be readily seen in Fig.~\ref{fig:N1000Densities}(d)-(f).
Notice the relatively small deviations between the predictions of the reduced model of Eq.~\eqref{eGPEsApprox2} and the exact eGPEs in the minority component density for such stronger attractions, where the former estimates a slightly smaller maximum density around $x=0$ and correspondingly slightly larger width (hardly visible in the scales of Fig.~\ref{fig:N1000Densities}(d)-(f)). 
This is attributed to the fact that the minority component density develops a Gaussian profile upon increasing $\abs{G_C}$, hence imprinting a progressively sharper density hump at the majority droplet distribution. 
Thus, the neglected energy terms, proportional e.g. to $\sim n_C^{5/2}/n$, become enhanced in spite of the large particle number imbalance among the droplet and component $C$. 
Therefore, despite the condition $N_C/N \ll 1$,  sufficiently strong intercomponent attraction can locally compensate for the population imbalance by substantially enhancing $n_C$ in the overlap region, thereby reducing the accuracy of the expansion in $N_C/N$.

\section{Summary and Perspectives}\label{sec:SummaryAndOutlook} 

We studied the stability and phases of 1D three-component bosonic mixtures composed of two symmetric species and a third, independent, component within the LHY framework. 
The two components with repulsive (attractive) intracomponent (intercomponent) interactions form a strongly self-bound symmetric  droplet which is coupled to the third species. 
To assess the composite phases of the three-component mixture, we computed the ground-state densities and their spatial overlap, the chemical potentials and the low-amplitude excitation spectrum (using BdG analysis) as a function of the atom number of the symmetric droplet or the intercomponent interaction between the symmetric droplet and the third component. 
Three main phases arise referring to an untrapped, partially trapped and fully trapped third component by the droplet, which do not share two-component counterparts.  

First, the particle balanced setup between the symmetric droplet and the third component was briefly addressed. 
In this case, the third component morphs into a spatially extended FT structure lying mainly outside the arguably more localized symmetric  droplet and being significantly perturbed by the presence of the latter featuring a pronounced density peak in the overlap region. 
The system is found to be (relatively) weakly bound, with the chemical potential increasing as the intercomponent attraction is reduced. 
Meanwhile, the low-amplitude collective modes gradually cross the particle-emission threshold from below as the intercomponent attraction increases. 

Focusing on the particle-imbalanced mixture within the weakly attractive regime, we find that upon increasing the atom number of the symmetric  droplet, the minority component becomes progressively fully trapped by the droplet, and eventually both species acquire comparable spatial extents. 
This transition is identified by the spatial overlap between the third component and the droplet, and it is accompanied by a decrease in the (negative) chemical potentials, indicating the formation of stronger self-bound configurations. 
This crossover is also clearly imprinted in the excitation spectrum. 
All modes (except the lowest-lying one) cross the particle-emission threshold from below as the system approaches the transition region and emerge again afterwards from above, featuring a characteristic pairing tendency in the trapped regime.

We also address the impact of the intercomponent coupling in the states of particle imbalanced mixtures considering variations within the attractive interaction MF stability regime to the decoupled limit. 
For strong attractions, the minority component assembles in a Gaussian configuration being fully trapped by the (two-component) droplet, which in turn displays localized undulations in its FT profile at the vicinity of the third component. 
For moderate attractions, the densities of all components transition into FT configurations of comparable spatial extent. 
Close to the decoupled limit, the minority component becomes loosely trapped and gradually delocalizes into an extended, low density FT droplet, reminiscent to the one of the particle balance case. 
As expected, the system becomes less self-bound with decreasing attraction, and the collective modes in the weakly interacting regime exhibit a transition, crossing the particle emission threshold, similarly to the behavior observed for different particle number. 
Interestingly, for stronger attractions the excitation spectrum possesses a complex interaction-dependent behavior with the ensuing modes experiencing multiple avoided crossings. 

To understand better the numerically identified droplet states, we derive different effective models based on a reduction of the LHY energy density in the limit of large particle imbalance between the majority droplet and the third component. 
It turns out that the reduced model accounting for the MF interactions and an effective potential to the minority species created by the majority droplet yields excellent agreement with the predictions of the complete LHY theory described by the full eGPEs. 
Other most simplified models, e.g. neglecting the MF coupling or based on the Thomas-Fermi approximation lead to noticeable deviations with the full system behavior.

Our results open several intriguing directions for future studies especially for particle imbalanced droplet configurations, while providing an initial overview of their ground-state phases and spectrum. 
Certainly, exploring the effect of an external harmonic trap in the discussed configurations represents an avenue of experimental interest. 
An immediate extension is to investigate the dynamical response of the particle imbalanced configurations following a quench or a sweep of the intercomponent coupling exploiting the observed avoided crossings between the collective modes across different localization regimes. 
Another interesting possibility is to explore the characteristics and properties of composite nonlinear excitations, as it was done for their single-component droplet counterpart~\cite{Katsimiga_solitons}, where the effective models developed herein may assist in extracting relevant analytical waveforms. 
Additionally, benchmarking the predictions of the LHY theory presented herein against ab-initio many-body methods~\cite{cao2017unified,ParisiGiorginiMonteCarlo,Mistakidis2021} aiming also to understand the underlying correlation patterns remains a completely open question for three-component mixtures. 
Finally, our setup can be used to investigate Bose polaron physics~\cite{BosePolaronDemler} in quantum droplets and to characterize, for instance, the lifetime, residue and induced interaction between the dressed minority atoms.

\section*{Acknowledgments}
This material is based upon work supported by, or in part by, the Army Research Laboratory and the Army Research Office under contract/grant number W911NF261A043 (E.G.C. and S.I.M.).
~S.I.M. thanks P.G. Kevrekidis, G.C. Katsimiga and G.A. Bougas for fruitful discussions on the topic of droplets.

\appendix
%\section*{APPENDIX}

\onecolumngrid
\section{Stability matrix of the three-component mixture}\label{app:SM}

In the main text, we have used BdG analysis to infer the stability and the behavior of the collective excitations of the three-component mixture comprising of two components forming a symmetric  droplet and a third component $C$. 
To do so, we have linearized the original eGPEs [Eq.~(\ref{eGPETwoPlusOneComponent})] as outlined in Section~\ref{sec:BdG_stab_matrix} to obtain the system of BdG equations containing the $4 \times 4$ stability matrix, dubbed $\mathcal{A}$. 
The elements of the latter, in dimensionless units, are given by
\begin{subequations}
\label{StabilityMatrixElements}
\begin{align}
\mathcal{A}_{11} = & - \frac{1}{2}\frac{\partial^2}{\partial x^2} - \mu + 2(g+g_{AB}) \abs{\Psi}^2 + G_C\abs{\Psi_C}^2 + \frac{1}{2}\Big(  \frac{\partial \mathcal{E}_{\textrm{LHY}}}{\partial n}  + \frac{\partial^2 \mathcal{E}_{\textrm{LHY}}}{\partial n^2} \abs{\Psi}^2 \Big), \\
\mathcal{A}_{12} = & (g+g_{AB} + \frac{1}{2} \frac{\partial^2 \mathcal{E}_{\textrm{LHY}}}{\partial n^2}) \Psi^2, \\
\mathcal{A}_{13} = & (G_C + \frac{1}{2} \frac{\partial^2 \mathcal{E}_{\textrm{LHY}}}{\partial n \partial n_C} ) \Psi \Psi_C^*, \\
\mathcal{A}_{14} = & ( G_C + \frac{1}{2} \frac{\partial^2 \mathcal{E}_{\textrm{LHY}}}{\partial n \partial n_C} ) \Psi \Psi_C, \\
\mathcal{A}_{21} = & (g+g_{AB} + \frac{1}{2} \frac{\partial^2 \mathcal{E}_{\textrm{LHY}}}{\partial n^2}) (\Psi^*)^2 = \mathcal{A}_{12}^*, \\
\mathcal{A}_{22} = & - \frac{1}{2}\frac{\partial^2}{\partial x^2} - \mu + 2(g+g_{AB}) \abs{\Psi}^2 + G_C\abs{\Psi_C}^2 + \frac{1}{2}\Big(  \frac{\partial \mathcal{E}_{\textrm{LHY}}}{\partial n}  + \frac{\partial^2 \mathcal{E}_{\textrm{LHY}}}{\partial n^2} \abs{\Psi}^2 \Big) = A_{11}, \\
\mathcal{A}_{23} = & (G_C + \frac{\partial^2 \mathcal{E}_{\textrm{LHY}}}{\partial n \partial n_C} ) \Psi^* \Psi_C^* = \mathcal{A}_{14}^*, \\
\mathcal{A}_{24} = & ( G_C + \frac{1}{2} \frac{\partial^2 \mathcal{E}_{\textrm{LHY}}}{\partial n \partial n_C} ) \Psi^* \Psi_C = \mathcal{A}_{13}^*, \\
\mathcal{A}_{31} = & ( 2G_C +  \frac{\partial^2 \mathcal{E}_{\textrm{LHY}}}{\partial n \partial n_C} ) \Psi^* \Psi_C,  \\
\mathcal{A}_{32} = & ( 2G_C +  \frac{\partial^2 \mathcal{E}_{\textrm{LHY}}}{\partial n \partial n_C} ) \Psi \Psi_C, \\
\mathcal{A}_{33} = & - \frac{1}{2}\frac{\partial^2}{\partial x^2} - \mu_C + 2 g_C  \abs{\Psi_C}^2 + 2G_C\abs{\Psi}^2 + \frac{\partial \mathcal{E}_{\textrm{LHY}}}{\partial n_C}  + \frac{\partial^2 \mathcal{E}_{\textrm{LHY}}}{\partial n_C^2} \abs{\Psi_C}^2,  \\
\mathcal{A}_{34} = & ( g_C + \frac{\partial^2 \mathcal{E}_{\textrm{LHY}}}{\partial n_C^2} ) \Psi_C^2, \\
\mathcal{A}_{41} = & ( 2 G_C + \frac{\partial^2 \mathcal{E}_{\textrm{LHY}}}{\partial n \partial n_C} ) \Psi^* \Psi_C^* = \mathcal{A}_{32}^*, \\
\mathcal{A}_{42} = & ( 2 G_C + \frac{\partial^2 \mathcal{E}_{\textrm{LHY}}}{\partial n \partial n_C} ) \Psi \Psi_C^* = \mathcal{A}_{31}^*, \\
\mathcal{A}_{43} = & ( g_C + \frac{\partial^2 \mathcal{E}_{\textrm{LHY}}}{\partial n_C^2} )  (\Psi_C^*)^2 = \mathcal{A}_{34}^*, \\
\mathcal{A}_{44} = & - \frac{1}{2}\frac{\partial^2}{\partial x^2} - \mu_C + 2g_C  \abs{\Psi_C}^2 + 2G_C\abs{\Psi}^2 + \frac{\partial \mathcal{E}_{\textrm{LHY}}}{\partial n_C}  + \frac{\partial^2 \mathcal{E}_{\textrm{LHY}}}{\partial n_C^2} \abs{\Psi_C}^2  = \mathcal{A}_{22}.
\end{align}
\end{subequations}

In the above expressions, the second derivatives with respect to the same densities take the form
\begin{subequations} 
\begin{align}
\label{LHY2ndDerivatives}
\frac{\partial^2 \mathcal{E}_{\textrm{LHY}}}{\partial n^2} =& - \frac{\sqrt{2m}}{4\pi h  } \Bigg(  \sqrt{2}  \frac{(g - g_{AB})^{\frac{3}{2}}}{ \sqrt{n}} + \frac{\Big( g + g_{AB} + \frac{\partial Q}{\partial n}  \Big)^2} {2\sqrt{W + Q}} + \frac{\Big( g + g_{AB} - \frac{\partial Q}{\partial n}  \Big)^2 }{2\sqrt{W - Q} } \\ \nonumber
&+ \frac{\partial^2 Q}{\partial n^2} \Big( \sqrt{W + Q} - \sqrt{W - Q}\Big) \Bigg), \\
\frac{\partial^2\mathcal{E}_{\textrm{LHY}}}{\partial n_C^2} =& - \frac{\sqrt{2m}}{4\pi h }  \Bigg(  \frac{\Big( g_C + \frac{\partial Q}{\partial n_C} \Big)^2} { 2\sqrt{W + Q}} +  \frac{\Big(  g_C - \frac{\partial Q}{\partial n_C}\Big)^2} {2\sqrt{W - Q}} + \frac{\partial^2 Q}{\partial n_C^2} \Big( \sqrt{W + Q} - \sqrt{W - Q}\Big)  \Bigg).
\end{align}
\end{subequations}
Here, the $W$, $Q$ parameters are the ones introduced in the main text [see also Eq.~(\ref{ELHY2Plus1Params})] participating in the full LHY energy density [Eq.~(\ref{ELHY2Plus1})] of the considered three-component mixture. 
The second derivatives of $Q$ with respect to the majority droplet density $n$ and the one of component $C$ read   
\begin{subequations}
\begin{align}
&\frac{\partial^2 Q}{\partial n^2} = \frac{(g+g_{AB})^2}{Q} - \frac{[(g+g_{AB})^2 n - (g g_C + g_{AB} g_C - 4 G_C^2) n_C]^2}{Q^3}= \frac{1}{Q}[(g+g_{AB})^2 - (\frac{\partial Q}{\partial n})^2],\\  
&\frac{\partial^2 Q}{\partial n_C^2} = \frac{g_C^2}{Q} - \frac{[g_C^2 n_C - (g g_C + g_{AB} g_C - 4 G_C^2) n]^2}{Q^3}= \frac{1}{Q}[(g_C)^2 - (\frac{\partial Q}{\partial n_C})^2].
\end{align}
\end{subequations}
Finally, the mixed derivative term of the LHY energy density appearing in the matrix elements of the linearized system of equations is found to be 
\begin{align}
\label{mixedDerivative}
\frac{\partial^2\mathcal{E}_{\textrm{LHY}}}{\partial n \partial n_C} = \frac{\partial^2\mathcal{E}_{\textrm{LHY}}}{\partial n_C \partial n} = &- \frac{\sqrt{2m}}{4\pi h }  \Bigg(  \frac{\Big( g_C + \frac{\partial Q}{\partial n_C} \Big)\Big( g + g_{AB} + \frac{\partial Q}{\partial n} \Big)} { 2\sqrt{W + Q}} +  \frac{\Big(  g_C - \frac{\partial Q}{\partial n_C}\Big) \Big( g + g_{AB} - \frac{\partial Q}{\partial n} \Big)} {2\sqrt{W - Q}} \\
&+ \frac{\partial^2 Q}{\partial n \partial n_C} \Big( \sqrt{W + Q} - \sqrt{W - Q}\Big)  \Bigg),  \nonumber
\end{align}
where the relevant expression of the mixed derivative of the $Q$ parameter becomes
\begin{subequations}
\begin{align}
&\frac{\partial^2 Q}{\partial n \partial n_C} = \frac{\partial^2 Q}{\partial n_C \partial n} =  \frac{1}{Q} [( 4 G_C^2 - g g_C - g_{AB} g_C) - \frac{\partial Q}{\partial n}\frac{\partial Q}{\partial n_C}],\\
&\frac{\partial Q}{\partial n} = \frac{ (g+g_{AB})^2 n - (g g_C + g_C g_{AB} - 4G_C^2) n_C}{Q},\\  
&\frac{\partial Q}{\partial n_C} = \frac{g_C^2 n_C - (g g_C + g_C g_{AB} - 4G_C^2) n}{Q}.
\end{align}
\end{subequations}

\section{LHY energy density for large particle imbalances}\label{app:limit}

Applying directly a Taylor expansion to Eq.~\eqref{ELHY2Plus1} in terms of $n_C/n\approx0$ (highly particle imbalance limit), results in a divergent second derivative $\partial^2 \mathcal{E}_{\rm LHY}/\partial (n_C/n)^2 \Big|_{0}$.
As such, we cannot rely on the direct Taylor expansion of the LHY energy density in the case of highly particle imbalance between the majority (two-component) droplet and component $C$. 
Moreover, inspecting the form of the LHY energy density given by Eq.~\eqref{ELHYRecast} below, it is clear that we should expect some terms with rational, but not integer, exponents to be included in our expansion, which cannot be captured by the direct Taylor expansion. 
Formally, this suggests that the expansion should be based on a Puiseux series~\cite{willis2008compute,walker1950algebraic}, instead of a Taylor series.
For our purposes, we simply proceed by applying the Taylor expansion in Eq.~\eqref{ELHY2Plus1} term-by-term as follows. 
First we recast the LHY energy density of Eq.~\eqref{ELHY2Plus1} into the following form  
\begin{equation}
\begin{split}
\label{ELHYRecast}
\mathcal{E}_{\textrm{LHY}} = - \frac{\sqrt{2 m}}{6\hbar\pi}(gn)^{3/2} \Bigg( & \sqrt{8} f_{-}^{\frac{3}{2}} + \bigg[ \big( f_{+} + \frac{g_C}{g} \frac{n_C}{n} \big) + \Big(f_{+}^2 - 2f_C \frac{n_C}{n} + \frac{g_C^2}{g^2} (\frac{n_C}{n})^2 \Big)^{1/2}\bigg]^{\frac{3}{2}} \\
&+ \bigg[ \big(f_{+} + \frac{g_C}{g}\frac{n_C}{n}) - \Big(f_{+}^2 - 2f_C \frac{n_C}{n} + \frac{g_C^2}{g^2} (\frac{n_C}{n})^2 \Big)^{1/2}\bigg]^{\frac{3}{2}}\Bigg),
\end{split}
\end{equation}
where $f_\pm = 1 \pm g_{AB}/g$, and $f_C= g_C/g + g_{AB} g_C/g^2 - (2G_C/g)^2$. 
Note that in relation to Eq.~\eqref{ELHY2Plus1} and Eq.~\eqref{ELHY2Plus1Params} the square root terms in Eq.~\eqref{ELHYRecast} correspond to $Q/gn$, the term inside the small parentheses to $W/gn$, and the expression inside the square brackets to $(W\pm Q)/gn$. 
Expanding the square root term appearing in Eq.~(\ref{ELHYRecast}) yields  
\begin{equation}
\begin{split}
\label{Qapprox}
\frac{Q}{gn} &= \Big(f_{+}^2 - 2f_C \frac{n_C}{n} + \frac{g_C^2}{g^2} (\frac{n_C}{n})^2 \Big)^{1/2} \approx f_{+} - \frac{f_C}{f_{+}} \frac{n_C}{n} 
+ \Big[\frac{g_C^2}{2g^2f_{+}}  - \frac{f_C^2}{2f_{+}^3}  \Big](\frac{n_C}{n} )^2 + \mathcal{O}\bigg((\frac{n_C}{n})^3\bigg).
\end{split}
\end{equation}
From this expression, it is clear that the term corresponding to $(W - Q)/(gn) $ becomes proportional to $n_C/n$. 
Hence, the term $\Big((W - Q)/(gn)\Big)^{3/2} $ gives rise to rational powers of $n_C/n$ in the form of $(n_C/n)^{\nu + 3/2 }$, with $\nu \in N^*$.
For this reason, as a next step, we factor out $n_C/n$ and expand the resulting expression in the brackets with respect to $n_C/n$, obtaining
\begin{subequations} 
\begin{align}
\label{WminusQapprox} 
\Big(\frac{W - Q}{gn}\Big)^{3/2} &\approx \Big[ \big( \frac{g_C}{g} + \frac{f_C}{f_{+}} \big)\frac{n_C}{n} - \Big[\frac{g_C^2}{2g^2f_{+}}  - \frac{f_C^2}{2f_{+}^3}  \Big](\frac{n_C}{n} )^2 \Big]^{3/2}  
= (\frac{n_C}{n} )^{3/2}  \Bigg[ \big( \frac{g_C}{g} + \frac{f_C}{f_{+}} \big) - \Big[\frac{g_C^2}{2g^2f_{+}}  - \frac{f_C^2}{2f_{+}^3}  \Big]\frac{n_C}{n}  \Bigg]^{3/2} \\ \nonumber 
&\approx \big( \frac{g_C}{g} + \frac{f_C}{f_{+}} \big)^{3/2} (\frac{n_C}{n} )^{3/2} - \frac{3}{2} \sqrt{\frac{g_C}{g} + \frac{f_C}{f_{+}}} \bigg( \frac{g_C^2}{2g^2f_{+}}  - \frac{f_C^2}{2f_{+}^3}  \bigg) (\frac{n_C}{n} )^{5/2}
 + \mathcal{O}\bigg((\frac{n_C}{n} )^{7/2}\bigg),\\ 
\label{WplusQapprox}
(\frac{W + Q}{gn})^{3/2} &\approx \Big[2f_{+}^2 +\big( \frac{g_C}{g} - \frac{f_C}{f_{+}} \big)\frac{n_C}{n} + \Big[\frac{g_C^2}{2g^2f_{+}}  - \frac{f_C^2}{2f_{+}^3}  \Big](\frac{n_C}{n} )^2 \Big]^{3/2} \\ \nonumber 
&\approx (2f_{+})^{3/2} + 
\frac{3\sqrt{2f_{+}}}{2} \big( \frac{g_C}{g} - \frac{f_C}{f_{+}} \big)\frac{n_C}{n} +\\ \nonumber
&\Bigg(  \frac{3\sqrt{2f_{+}}}{2} \Big[\frac{g_C^2}{2g^2f_{+}}  - \frac{f_C^2}{2f_{+}^3}  \Big]  + 
\frac{3}{8\sqrt{2f_{+}}} \Big[\frac{g_C}{g}  - \frac{f_C}{f_{+}}  \Big]^2 \Bigg)   (\frac{n_C}{n} )^2
 + \mathcal{O}\bigg((\frac{n_C}{n})^3\bigg). 
\end{align}
\end{subequations} 
Substituting all the above terms into the LHY energy density of Eq.~(\ref{ELHYRecast}) and setting $f_G= (2\frac{G_C}{g})^2/f_{+}$ results in 
\begin{equation}
\label{ApproxE_LHY_steps}
\begin{split}
\mathcal{E}_{\rm LHY} \approx  
-\frac{2\sqrt{mg^3}}{3\hbar\pi} \Bigg[ 
(f_{-}^{3/2} + f_{+}^{3/2}) n^{3/2} 
+ \big( \frac{g_C}{g} - \frac{f_G}{2} \big)^{3/2}  n_C^{3/2} 
+ \frac{3\sqrt{f_{+}}}{4} f_G \sqrt{n}n_C 
+ \mathcal{O}(\frac{1}{\sqrt{N}})
\Bigg].      
\end{split}    
\end{equation}

\begin{figure*}[ht]
\centering   
\includegraphics[width=0.95\linewidth]{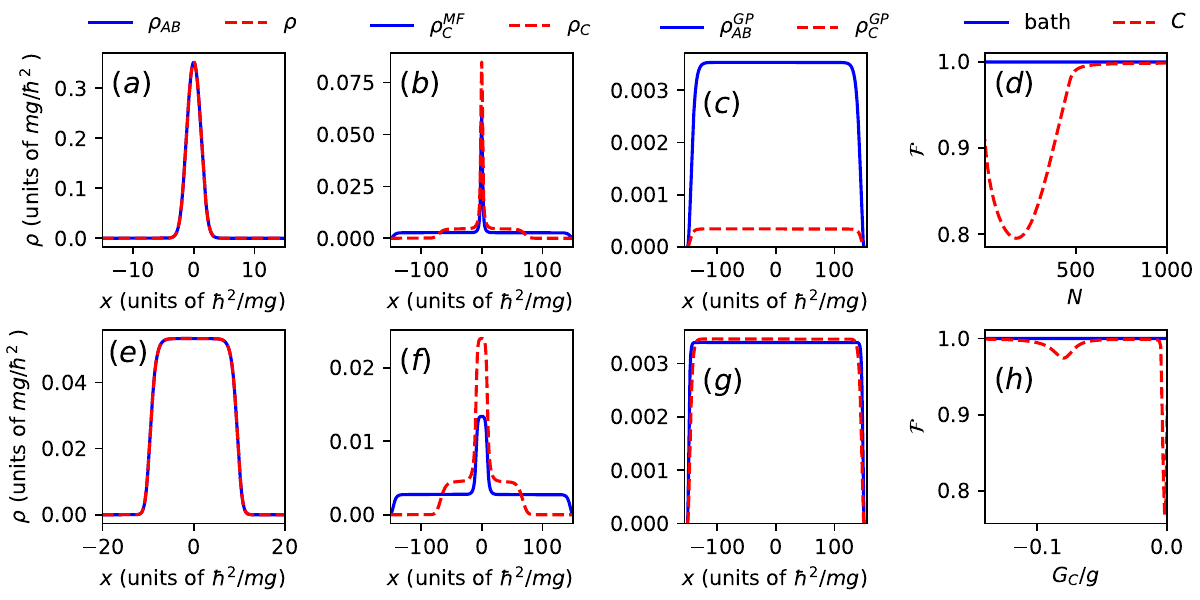}
\caption{Ground state densities of (a), (e) the two-component majority droplet and (b), (f) component $C$ containing $N_C=10$ bosons, with $g_C=g$, $g_{AB}=-0.96g$ using the full eGPE treatment [Eqs.~(\ref{eGPETwoPlusOneComponent})] (dashed lines) and the reduced model of Eqs.~(\ref{eGPEDropletPlusMF}) neglecting the LHY related to the minority species (solid lines).  
The system is characterized by (a), (b), (c) relatively small particle difference with $N=100$ and  $G_C=-0.01g$, or (e), (f), (g) weak intercomponent attraction $G_C=-0.001g$ and large population imbalance with $N=10^3$. 
In both cases, the minority component $C$ (panels (b), (f)) exhibits a self-bound behavior under the impact of the LHY term of the third component (dashed lines), while in its absence the minority species distribution has delocalized tails extending throughout the box potential and a smaller amplitude of the central peak structure (solid lines). 
Instead, the majority droplet (panels (a), (e)) is almost the same within both approaches.
Fidelity of the bath ($\mathcal{F}_{AB}$) and component $C$ ($\mathcal{F}_{C}$) wavefunctions (see text) for (d) varying atom number of the two-component droplet using  $N_C=10$, and $G_C=-0.01g$ as well as for (h) different intercomponent couplings $G_C$ when $N=10^3$, and  $N_C=10$. 
The fidelity of component $C$ is reduced signifying deviations between the two methods especially for different atom numbers in the two-component droplet (panel (c)). 
These differences originate from the absence of the LHY term of component $C$. 
In contrast $\mathcal{F}_{AB} \approx 1$ since the LHY contribution of the two-component droplet is taken into account in both approaches. 
Neglecting all LHY contributions, i.e. operating in the pure MF approximation (panels (c), (g)) yields delocalized structures in both subsystems. 
In all cases, the remaining interaction parameters correspond to $g_C=g$, and $g_{AB}=-0.96g$. 
}
\label{fig:MFcomparison}
\end{figure*}

\section{Stabilization role of the LHY contributions}\label{app:MF}

To highlight the impact of the quantum fluctuations of the genuine three-component system, as captured by the LHY energy of Eq.~\eqref{ELHY2Plus1}, we compare the predictions of the full eGPE framework of Eqs.~(\ref{eGPETwoPlusOneComponent}) used in the main text with an intermediate model.
The latter consists of a two-component droplet~\cite{PetrovLowD,Englezos2025Droplets} and a third minority species  featuring solely MF coupling, i.e. ignoring the LHY contribution appearing in the Eq.~(\ref{eGPETwoPlusOneComponent_b}) of component $C$. 
This investigation will allow us to appreciate the role of quantum fluctuations in the third species but also between the latter and the majority component. 
The corresponding coupled set of equations of motion of the aforementioned reduced model reads 
\begin{subequations}
\label{eGPEDropletPlusMF}
\begin{align}
\label{eGPEDropletPlusMF_a}
i\hbar\frac{\partial \psi_{\rm AB}}{\partial t}& = - \frac{\hbar^2}{2m} \frac{\partial^2 \psi_{\rm AB}}{\partial x^2} +\Big( (g+g_{AB}) \abs{\psi_{\rm AB}}^2 + G_C \abs{\psi^{\rm MF}_C}^2 -\frac{\sqrt{mg^3}(f_{-}^{3/2}+f_{+}^{3/2})}{2\hbar\pi}\abs{\psi_{\rm AB}} \Big)\psi_{\rm AB},  \\ 
i\hbar\frac{\partial \psi^{\rm MF}_C}{\partial t}& = - \frac{\hbar^2}{2m} \frac{\partial^2 \psi^{\rm MF}_C}{\partial x^2} + \Big( g_{C}\abs{\psi^{\rm MF}_{C}}^2 + 2G_C \abs{\psi_{\rm AB}}^2  \Big)\psi^{\rm MF}_C. \label{eGPEDropletPlusMF_b}
\end{align}
\end{subequations}
Here, $\psi_A=\psi_B\equiv \psi_{\rm AB}$ is the common wavefunction of the two symmetric bath components, and $\psi_C^{\rm MF}$ represents the wavefunction of the minority component. 
Our analysis below, is based on the density distributions of the involved components obtained within the different approaches and also the fidelity $\mathcal{F}_{AB}=\abs{ \braket{<\psi_{\rm AB} | \psi }}$ ($\mathcal{F}_C=\abs{ \braket{<\psi_C^{\rm MF} | \psi_C }}$) providing a quantitative measure of the deviation of the wave functions of the bath (minority species) between the two methods. 
These comparisons enable a concrete identification of the role of the quantum fluctuations in the third species.

As a first characteristic example illustrating the crucial effect of the third component LHY term, we invoke the system with $N=100$ particles in the majority two-component droplet, $N_C=10$ in the minority species and relatively weak intercomponent minority-droplet coupling $G_C=-0.01$. 
The resultant density distributions of the distinct subsystems are illustrated in Fig.~\ref{fig:MFcomparison}(a),(b) within both methods. 
As it can be clearly seen, the MF coupling alone in component $C$ fails to capture the self-bound character of the minority component [Fig.~\ref{fig:MFcomparison}(b)] as predicted by the full set of eGPEs described by Eqs.~\eqref{eGPETwoPlusOneComponent}.  
This becomes evident by the fact that in the absence of the LHY term of the third component [Eqs.~(\ref{eGPEDropletPlusMF})] the amplitude and width of the central structure in the density of the minority species are underestimated and most notably the flat tails of the distribution are extended throughout the box potential. 
Such a tendency is reminiscent of a gaseous behavior and it is in sharp contrast to the self-bound character predicted by the full set of eGPEs [Eqs.~(\ref{eGPETwoPlusOneComponent})]. 
On the other hand, we observe excellent agreement on the density configuration of the two-component droplet obtained with the different methods, see Fig.~\ref{fig:MFcomparison}(a). 
This is expected since in both models the LHY contribution of the two-component droplet is consistently taken into account. 
Similar deviations between the two approaches occur for different interactions as well, see e.g. Fig.~\ref{fig:MFcomparison}(e),(f) for a setting with $N=1000$, $N_C=10$ and $G_C=-0.001$. 
Apparently, the self-bound behavior of the minority component distribution is evident using the full eGPEs, while it is entirely absent when the LHY term of component $C$ is neglected  [Fig.~\ref{fig:MFcomparison}(f)]. 
Additionally, the amplitude of the central peak is lower without the LHY term. 
In contrast, the FT droplet of the majority components is the same between the two models, see Fig.~\ref{fig:MFcomparison}(d). Therefore, according to the above, we can infer that the LHY term of the minority component facilitates the stabilization of the latter.

To gain an overview of the importance of the LHY term of the third component, we next compute $\mathcal{F}_{AB}$ and $\mathcal{F_C}$ for different atom numbers in the two-component droplet [Fig.~\ref{fig:MFcomparison}(d)] while keeping fixed all other system parameters. 
It is found that the deviations between the two methods discussed above in capturing the self-bound minority species configurations are imprinted in the associated fidelity where $\mathcal{F}_C<1$ for a wide range of particle numbers in the droplet, until the particle imbalance becomes quite extreme, namely $N/N_C > 50$. 
However, similar deviations in $\mathcal{F}_C$ between the two approaches appear also for such extreme particle imbalances but for weaker attractions.  
This argument is supported by Fig.~\ref{fig:MFcomparison}(h) presenting the fidelities for varying attractive couplings between the two-component droplet and the impurity $G_C$.
Indeed, for weak couplings ($-0.005<G_C<0$) it is observed that $\mathcal{F_C} \ll 1$. 
Interestingly, there is an additional interaction regime in the vicinity of $G_C \approx-0.08$ where $\mathcal{F_C}<1$. 
This manifests the occurrence of deviations in the wave function of the minority species as predicted by the two models.  

Notably, this interaction region coincides with the one in which a series of avoided crossings, between the low-lying modes and the higher-lying modes which sharply descend from above the particle emission threshold, was identified in the excitation spectrum of the three-component mixture (see also Fig.~\ref{fig:N1000Energies}(b)). 
According to our analysis, these avoided crossings do not take place in the simplified model where the LHY contribution of the third component is neglected. 
We remark that the non self-bound character of the system within the model of Eqs.~(\ref{eGPEDropletPlusMF}) can be independently inferred by evaluating the chemical potential of the minority species which turns out to be positive, i.e. $\mu_C^{\rm MF}>0$,  in all cases where component $C$ is not fully immersed in the two-component droplet (not shown for brevity).
In contrast to the above, the fidelity of the bath components among the two methods is approximately unity for the considered parametric variations, see Fig.~\ref{fig:MFcomparison}(d), (h). 
Let us note, however, that the LHY term of the two-component majority subsystem is crucial for capturing the self-bound droplet. 
Indeed, relying on the standard MF Gross-Pitaevskii (GP) equations~\cite{pethick2008bose}, where all LHY contributions are not taken into account leads to a delocalized density profile extending over the entire box length. 
This is explicitly shown in Fig.~\ref{fig:MFcomparison}(c) for $N=100$ and $G=-0.01$ and in Fig.~\ref{fig:MFcomparison}(g) for $N=1000$ and $G=-0.001$, indicated by $\rho_{AB}^{GP}$ and $\rho_{C}^{GP}$ for the majority and minority components respectively. 

It is important to emphasize that these density profiles predicted within the standard MF approach for atomic mixtures are radically different from the self-bound droplet-like structures discussed in the main text.
In particular, these MF density configurations crucially depend on the box length in contrast to the more accurate eGPE results which are insensitive to the box length, provided that it is larger than the inherent spatial extent of the self-bound configurations. 
Moreover, within the MF approximation, although in some cases the minority component can be trapped by the majority one and feature negative chemical potential, the majority component is always characterized by a positive chemical potential.
This clearly indicates that the MF configurations can only remain stable for a given box length, while they would immediately disperse upon release, in sharp contrast to the stationary self-bound droplet solutions.
Summarizing, the above analysis underscores the key role of each LHY term in stabilizing self-bound states in both the minority and the two-component majority species. 

\section{Interaction dependence on the ground state energy of the three-component system}\label{app:phase_diagram}

\begin{figure*}[ht]
\centering   
\includegraphics[width=0.95\linewidth]{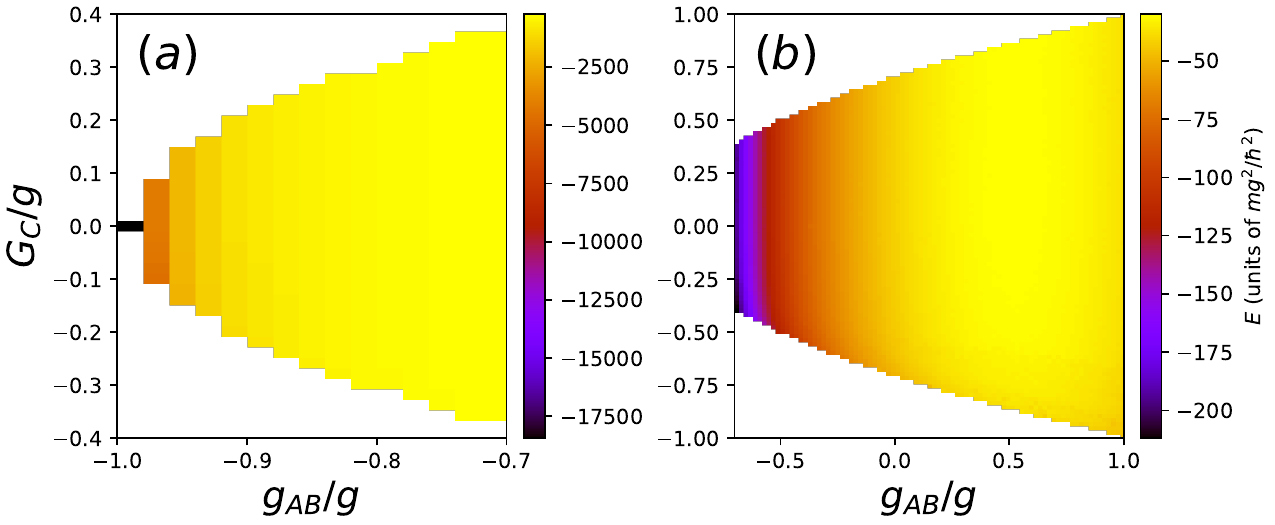}
\caption{Ground state energy of the three-component particle imbalanced setting throughout the MF stability regime. For better visualization we illustrate the interaction regions (a) $-g\leq g_{AB} \leq -0.7g$ and (b)  $-0.7g \leq g_{AB} \leq g$ separately. 
Notice that the total energy is negative for all values within the MF stability regime, evincing the self-bound nature of the system. 
The white areas in both panels designate the parametric regions where the LHY term becomes imaginary. 
In all cases, the remaining interaction parameters correspond to $g_C=g$, and we consider $N=10^3$ atoms in each of the two identical components $A$ and $B$ and $N_C=10$ bosons in component C. }
\label{fig:Phase_diagram}
\end{figure*}

To get an overview on the behavior of the ground state of the particle imbalanced system featuring $N=10^3$ atoms in the two identical components $A$ and $B$ and $N_C=10$ atoms in component C, we calculate its ground state energy for varying intercomponent interactions $g_{AB}$ and $G_C$ and fixed $g_C=g$.
Namely, the energy of the underlying three-component system is defined as:
\begin{equation}
E = \int \left[ \frac{\hbar^2}{m}\abs{\frac{\partial \psi}{\partial x}} + \frac{\hbar^2}{2m}\abs{\frac{\partial \psi_C}{\partial x}} + (g+g_{AB})\abs{\psi}^4 + \frac{g_C}{2}\abs{\psi_C}^4 + 2G_C \abs{\psi}^2\abs{\psi_C}^2 + \mathcal{E}_{LHY}\right]dx,
\end{equation}
where $\psi$, $\psi_C$ represent the ground state wavefunction of the two identical components and component $C$, respectively.
As it can be seen in Fig.~\ref{fig:Phase_diagram}, the ground state energy of the system remains negative throughout the MF stability regime, confirming its self-bound nature. 
Evidently, the intercomponent interaction strength between the two identical components $g_{AB}$ has the dominant effect on the ground state energy, owing also to the much larger particle number occupying these two components.
However, close inspection of the energy indicates that it also slightly increases as the intercomponent interaction strength $G_C$ is tuned from negative to positive values (hardly visible at this scale). 
Moreover, as anticipated the total energy of the system becomes more negative for stronger attractive interspecies interactions ($g_{AB}$) among the majority components, compare Fig.~\ref{fig:Phase_diagram}(a) and (b). 
Finally, as shown in Fig.~\ref{fig:Phase_diagram}(a) the ground state energy sharply decreases as $g_{AB} \xrightarrow{} -g$, evincing the presence of strongly self-bound configurations.

\twocolumngrid

\bibliographystyle{apsrev4-1}
\bibliography{ref_drops}

\end{document}